\documentclass[aps,preprint,amssymb,superscriptaddress]{revtex4-1}
\usepackage{floatrow}
\newfloatcommand{capbtabbox}{table}[\FBwidth]
\usepackage{graphicx}
\usepackage{dcolumn}
\usepackage{bm}
\usepackage{natbib}
\usepackage{CJK}
\usepackage[utf8]{inputenc}
\usepackage[T1]{fontenc}
\usepackage{mathptmx}
\usepackage{gensymb}
\usepackage{graphicx}
\usepackage{epstopdf}
\usepackage{dcolumn}
\usepackage{bm}
\usepackage{amsmath}
\usepackage{mathtools}
\usepackage{relsize}
\usepackage{multirow}
\usepackage[colorlinks]{hyperref}
\begin{document}


\title{Calculating the Hyperfine Tensors for Group-IV Impurity-Vacancy Centers in Diamond: A Hybrid Density- Functional Theory Approach}

\author{Rodrick Kuate Defo}
\affiliation{John A. Paulson School of Engineering and Applied Sciences, Harvard University, Cambridge, MA 02138}
\affiliation{Department of Electrical and Computer Engineering, Princeton University, Princeton, NJ 08540}
\author{Efthimios Kaxiras} 
\affiliation{John A. Paulson School of Engineering and Applied Sciences, Harvard University, Cambridge, MA 02138}
\affiliation{Department of Physics, Harvard University, Cambridge, MA 02138}
\author{Steven L. Richardson} 
\affiliation{John A. Paulson School of Engineering and Applied Sciences, Harvard University, Cambridge, MA 02138}
\affiliation{Department of Electrical and Computer Engineering, Howard University, Washington, DC 20059}
\date{\today}

\begin{abstract}

The hyperfine interaction is an important probe for understanding the structure and symmetry of defects in a semiconductor. Density-functional theory has shown that it can provide useful first-principles predictions for both the hyperfine tensor and the hyperfine constants that arise from it. Recently there has been great interest in using group-IV impurity-vacancy color centers  X$V^-$ (where X = Si, Ge, Sn, or Pb and $V$ is a carbon vacancy) for important applications in quantum computing and quantum information science. In this paper, we have calculated the hyperfine tensors for these  X$V^-$ color centers using the HSE06 screened Hartree-Fock hybrid exchange-correlation functional with the inclusion of core electron spin polarization. We have compared our results to calculations which only use the PBE exchange-correlation functional without the inclusion of core electron spin polarization and we have found our results are in very good agreement with available experimental results. Finally, we have theoretically shown that these X$V^-$ color centers exhibit a Jahn-Teller distortion which explains the observed anisotropic distribution of the hyperfine constants among the neighboring $^{13}$C nuclear spins. 

\end{abstract}

\maketitle


\section{\label{intro}INTRODUCTION}
The hyperfine interaction is the magnetic interaction between the magnetic moment of a nucleus and the magnetic moment of an electron. This interaction is important in a number of applications including mapping the sky through the presence of hydrogen molecules~\cite{Purcell1951,*Muller1951}, studying conduction electrons through the Knight shift~\cite{kittel2004introduction}, and exploring the electrostatic hyperfine tuning of phosphorus donors in silicon.~\cite{Dreher} For the case of the contribution from the state of zero angular momentum of the electron, the hyperfine interaction has two major terms: a contribution from the state of zero angular momentum of the electron which is the so-called (isotropic) Fermi-contact term and the dominant contribution due to the (anisotropic) magnetic dipole-dipole interaction. Explicitly, it has the form,~\cite{griffiths1982hyperfine,*griffiths2014introduction,*slichter2013principles}

\begin{equation}
 \frac{2\mu_0}{3}g_e\mu_eg_J\mu_J\mathbf{\hat{S}}^J\cdot\mathbf{\hat{S}}^e\delta(\mathbf{R}_J)+\frac{1}{4\pi}\mu_0g_e\mu_eg_J\mu_J\frac{1}{r^3}\left[3(\mathbf{\hat{S}}^J\cdot\hat{r})(\mathbf{\hat{S}}^e\cdot\hat{r})-\mathbf{\hat{S}}^J\cdot\mathbf{\hat{S}}^e\right],
 \label{eq:interaction}
\end{equation}
where $\mathbf{\hat{S}}^J$ is the nuclear spin operator for a nucleus at position $\mathbf{R}_J$, $\mathbf{\hat{S}}^e$ is the electron spin operator for an electron at a distance $r$ in the direction $\hat{r}$ relative to the nucleus $J$, $\mu_0$ is the permeability of vacuum, $\mu_J$ is the nuclear magneton of nucleus $J$, $\mu_e$ is the Bohr magneton, and $g_J$ and $g_e$ are the g-factors corresponding to the nucleus and electron, respectively. Note that in Eq. (\ref{eq:interaction}), we have taken $\hbar = 1$. 

We can then write the Hamiltonian that captures this interaction in terms of the hyperfine tensor $\mathbf{A}$,
\begin{equation}
    \mathcal{H} = \mathbf{\hat{S}}^J\cdot\mathbf{A}\cdot\mathbf{\hat{S}}^e.
    \label{eq:Hamiltonian}
\end{equation}
For numerical convergence, we replace $\delta(\mathbf{r})$ by  $\delta_T(\mathbf{r})$, where the new delta function $\delta_T(\mathbf{r})$ is localized within the Thomson radius, $r_T = Ze^2/(4\pi\epsilon_0m_ec^2).$~\cite{Blochl2000} In constructing Eq. (\ref{eq:Hamiltonian}) from Eq. (\ref{eq:interaction}), we integrate over the electronic wavefunction, which introduces the spin density $\sigma(\mathbf{r})$. Thus, the hyperfine tensor has the form~\cite{Blochl2000}
\begin{equation}
    A_{ij}^{(J)} = \mu_0\gamma_e\gamma_J\left[\frac{2}{3}\delta_{ij}\int \delta_T(\mathbf{r-R}_J)\sigma(\mathbf{r})d\mathbf{r}+\frac{1}{4\pi}W_{ij}(\mathbf{R}_J)\right],
    \label{eq:Atensor}
\end{equation}
where the first term in the square brackets is the isotropic Fermi-contact term originating from electrons in spherically symmetric orbitals and the second term in the square bracket
\begin{equation}
    W_{ij}(\mathbf{R}) = \int\left[\frac{3(\mathbf{r-R})_i(\mathbf{r-R})_j}{|\mathbf{r-R}|^5}-\frac{\delta_{ij}}{|\mathbf{r-R}|^3}\right]\sigma(\mathbf{r})d\mathbf{r}
\end{equation}
is the anisotropic magnetic dipole-dipole contribution. In Eq. (\ref{eq:Atensor}) we have defined the gyromagnetic ratio of nucleus $J$ as $\gamma_J = g_J\mu_J/\hbar$ and  the gyromagnetic ratio of the electron $e$ as $\gamma_e = g_e\mu_e/\hbar$. Following the convention of Gali \textit{et al.}~\cite{Gali4} we can decompose the hyperfine tensor $A^{(J)}_{ij}$ into its principal values $A_{xx}$, $A_{yy}$, and $A_{zz}$ and we refer to these as the hyperfine constants. The hyperfine interaction causes splittings between electronic energy levels and between nuclear energy levels. With the application of an external magnetic field, these energy levels are split and energy transitions among these states can be measured.~\cite{kittel2004introduction,Galvan} Theory can then be used to relate such energies to the hyperfine tensor above.~\cite{Galvan}

The hyperfine interaction has some important consequences when applied to the case of a paramagnetic defect in a semiconductor. In particular, these hyperfine constants can yield important information about which types of atoms comprise the defect, where the atoms that surround the defect are located, what the overall symmetry of the defect is, and upon which atoms the electron spin density $\sigma(\mathbf{r})$ is primarily located.~\cite{Walle, Szasz} In addition, first-principles calculations for these hyperfine constants can be directly compared with experimental results for defects in semiconductors.~\cite{Walle, Szasz}
The spin-based qubits investigated by Childress \textit{et al.}\,(N$V^-$)~\cite{Childress281} and Dreher \textit{et al.}\,(P)~\cite{Dreher} are paramagnetic defects and are often solid state single photon emitters (SPEs) consisting of impurity-vacancy defects and they have been widely studied. As far as practical applications are concerned, they show promise in quantum computing and quantum information processing in addition to the more traditional field of metrology. 
~\cite{Doherty2013nitrogen,*awschalom2018quant,*DEFO2019,*Kuate,*Kuate8,*Ladd2010,*Pfaff532} 

As we have discussed before, one important example of a paramagnetic defect in a semiconductor is the negatively-charged N$V$ center in diamond. Previous attempts to calculate the hyperfine tensor for the N$V^-$ have used the local spin density approximation of Ceperley-Alder~\cite{ceperley1980ground} in density-functional theory (DFT), as parametrized by Perdew and Zunger~\cite{perdew1981self},  with a 512-atom supercell in an all-electron PAW calculation~\cite{Gali4, GaliPBE} and the Perdew-Burke-Ernzerhof (PBE)~\cite{Perdew} exchange-correlation functional of DFT with a 512-atom supercell in an all-electron calculation.~\cite{Smeltzer_2011,Gali2} Unfortunately, all-electron calculations are extremely computationally demanding particularly for large supercell sizes. While calculations involving \textit{ ab initio} pseudopotentials are far less expensive than all-electron calculations, they inherently disregard the contribution of the spin polarization of core electrons to the isotropic hyperfine constants which are already included by construction in all-electron calculations. Yazyev \textit{et al.}~\cite{Yazyev} have proposed a scheme to circumvent this problem by including the contribution of core electron spin polarization to the hyperfine interaction within pseudopotential electronic structure methods and this procedure is now implemented in VASP~\cite{Kresse1,*Kresse2,*Kresse3} as of version 5.3.2. The accuracy of this method has been demonstrated by comparison with all-electron calculations.~\cite{Yazyev} 

Szasz \textit{et al.}~\cite{Szasz} have subsequently used the formalism of Yazyev \textit{et al.}~\cite{Yazyev} with both the PBE and the HSE06 screened Hartree-Fock hybrid exchange-correlation functional~\cite{Heyd, Krukau} to compute the hyperfine constants for the N$V^-$ at the nearest-neighbor $^{13}$C nuclear spins closest to the N atom. They found good agreement for their HSE06 results with the experiments of Felton \textit{et al.}~\cite{Felton} by within at most a few percent. It is interesting to note that they also found good agreement for their PBE results without the inclusion of core electron spin polarization. These authors explained their result as being fortuitous as it likely resulted from a cancellation of the error in underestimating the localization of the defect wavefunctions and the error in neglecting core electron spin polarization.~\cite{Szasz} 

\begin{figure}[ht!] 

\centering
\begin{minipage}[c]{0.4\textwidth}\centering\includegraphics[width=0.7\textwidth]{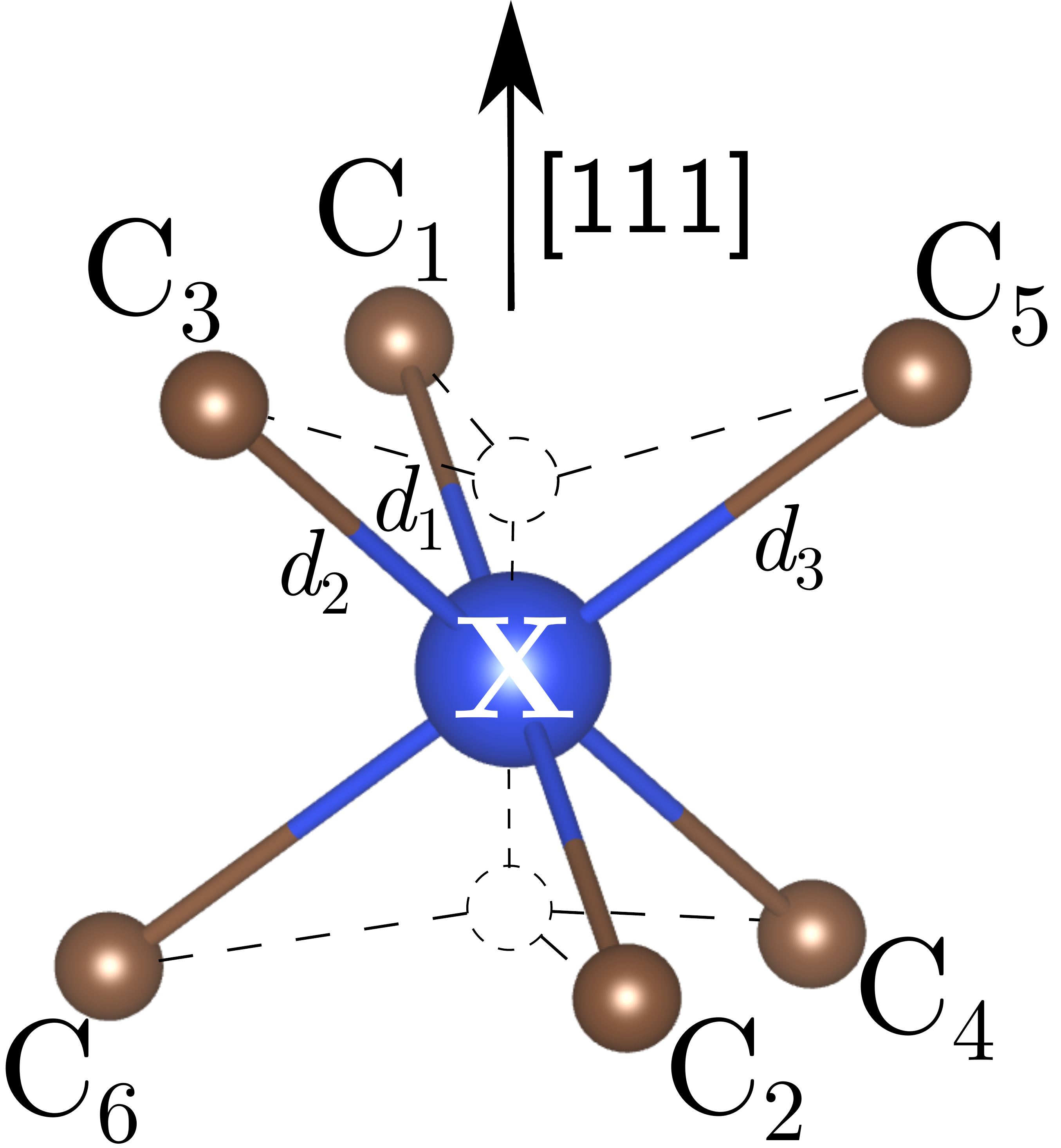}\end{minipage}\parbox{0.4\textwidth}{\vspace{15pt}\begin{tabular}{|c|l|l|l|l|}
\hline\hline 
 & Si$V^-$ & Ge$V^-$ & Sn$V^-$ & Pb$V^-$ \\
\hline 
$d$ &  ~1.96$^{\rm a}$ & ~2.01$^{\rm a}$ & ~2.08$^{\rm a}$ & ~2.12$^{\rm a}$   \\
$d_1$ &  ~1.96  & ~2.02  & ~2.09 &  ~2.13 \\
$d_{2,3}$ & ~1.96 & ~2.00 & ~2.08 & ~2.13 \\
\hline
\end{tabular}}
\centering

\label{fig:structure}
\caption{Structure of a group-IV impurity-vacancy color center (X$V^-$ where X = Si, Ge, Sn, or Pb and $V$ is a carbon vacancy) from the perspective along the [111] direction. The nearest-neighbor carbon atoms are numbered and shown in brown while the dashed circles are carbon vacancies. Bond lengths $d_1$, $d_2$ and $d_3$ between carbon atoms and X atom are in \AA. The undistorted bond lengths ($d$) from structural relaxations resulting in all nearest-neighbor carbon atoms being located at the same distance from the X atom are from other theory and are also expressed in~\AA. $^{\rm a}$ Ref. \cite{Gali3}.}
\end{figure}

While the N$V^-$ color center is certainly the most promising impurity-vacancy color center because it functions at room temperature, only 4 \% of its fluorescence is found in the zero-phonon line (ZPL). Another issue with the N$V^-$ is that it lacks inversion symmetry and thus possesses an electric dipole moment which makes it quite susceptible to external noise and local fields. Both of these effects tend to broaden the lineshape of excitation transitions thus causing serious problems for its viability as a solid state single photon emitter.~\cite{Deak}
One way to avoid the deleterious effects of the lack of inversion symmetry in N$V^-$ is to look at SPEs that possess inversion symmetry such as group-IV impurity-vacancy centers (X$V^-$) (where X= Si, Ge, Sn, or Pb and $V$ is a carbon vacancy) in the diamond structure. These color centers consist of an interstitial X atom placed in between two adjacent missing carbon atoms with an extra electron added to form a negative charge state. (\textit{Cf.} Fig. \ref{fig:structure})
There has been a considerable amount of both theoretical and experimental work on X$V^-$ impurity-vacancy
centers (Si~\cite{neu2011single,*haussler2017photoluminescence}, Ge~\cite{Ekimov,*Iwasaki,*Ralchenko,*Ekimov2,*Palyanov,*Palyanov2015germanium,*Bhaskar,*bray2018single}, Sn~\cite{Iwasaki2,*ekimov2018tin, *tchernij2017single,*palyanov2019high, *Alkahtani2018tin, *rugar2019char, *wahl2020direct,*fukuta2021sn}, Pb\cite{Trusheim, *tchernij2018single} in diamond and combination studies of several of these  centers~\cite{ekimov2019effect,*haussler2017photoluminescence} ) to produce them experimentally both as isolated species and in hybrid structures, to measure their optoelectronic and spin properties, and to enhance their spin coherence times by implanting them in cantilevers.
Childress \textit{et al.}~\cite{Childress281} have shown the importance of the hyperfine interaction regarding spin coherence times of N$V^-$ color centers in diamond, suggesting that a density-functional theory (DFT) investigation of the hyperfine interaction for X$V^-$ color centers in diamond could also lead to useful insights and predictions regarding spin coherence.

Many authors have performed experimental and theoretical studies of the hyperfine constants for the neutral Si$V^0$ impurity-vacancy center in diamond which is known to be paramagnetic. Gali and Maze~\cite{gali2013ab} have computed the hyperfine constants of the nearest-neighbor $^{13}$C nuclear spins for Si$V^0$ at the HSE06 level of theory and found agreement with the experimental results of Edmonds and \textit{et al.}~\cite{edmonds2008electron} to within at most a few percent. Their calculation included the effect of core electron spin polarization. They also computed the hyperfine constants for Si$V^0$ at the $^{29}$Si isotope atom of silicon, but found their result to be in error with experiment by roughly 20\%. We find similar disagreement in our own calculations for Si$V^0$ at the $^{29}$Si isotope atom of silicon. It is interesting to note that the method of Yazyev \textit{et al.}~\cite{Yazyev} was compared with all-electron calculations for systems containing H, C, N, O and F, but not Si, which may explain the disagreement. As we restrict ourselves in this work to the investigation of the hyperfine interaction with only carbon nuclear spins, our ensuing results should not suffer from the same inaccuracy when compared with experiment.

Encouraged by the success of Gali and Maze~\cite{gali2013ab} in using HSE06 to calculate the hyperfine constants at the nearest-neighbor $^{13}$C nuclear spins of Si$V^0$ using core electron spin polarization, we report in this paper the hyperfine constants for the other important group-IV impurity-vacancy centers (Ge$V^-$, Sn$V^-$, and Pb$V^-$) calculated at the same level of theory. This information is very important because of the recent attention that these defects have attracted in the field of quantum information science. We have also computed the electron spin densities for these centers and used this information to explain the trends we observed for the hyperfine constants of the nearest-neighbor $^{13}$C nuclear spins which surround the defect. While experimental data for these impurity-vacancy centers are currently unavailable, we are encouraged that that our results serve as a control study since they are also in good agreement with the calculations of Gali and Maze~\cite{gali2013ab} for the case of Si$V^0$ and should be helpful as a guide for future experimental studies. We have argued that the Jahn-Teller distortion exhibited at low temperatures by group-IV impurity-vacancy color centers can be exploited to enhance the spin coherence time of those color centers. Specifically, this distortion leads to $^{13}$C nuclear spins along a certain axis coupling more strongly than nuclear spins in other directions to the electrons associated with the X$V^-$ color center through the hyperfine interaction. This result suggests that several iterations of reheating and cooling samples may allow the system to find a distortion that gives an optimal coupling between the $^{13}$C nuclear spins and the electrons associated with the X$V^-$ color center. We also argue that this anisotropy will lead to a direction-dependent spin diffusion barrier that should be taken into account in constructing hybrid systems of color centers. We first review and discuss our computational methods in Section \ref{methods}. In Section \ref{discussion} we present our results and discussion and in Section \ref{diffusivity} we provide the conclusions of our paper.

\section{\label{methods}COMPUTATIONAL METHODS}

We performed first-principles DFT calculations for the various defect structures using the VASP code.~\cite{Kresse1,Kresse2,Kresse3}
For the exchange-correlation energy of electrons we used both the generalized gradient approximation (GGA) as parametrized by (PBE)~\cite{Perdew} and HSE06 with the original parameters.~\cite{Heyd,Krukau} Scalar relativistic effects~\cite{Perdew} were included with a projector augmented wave (PAW) pseudopotential.~\cite{Kresse3}
For the stoichiometric conventional unit cell the atomic positions were relaxed until the magnitude of the Hellmann-Feynman forces was smaller than $10^{-4}$ eV$\cdot$\AA$^{-1}$~on each atom and the lattice parameters were concurrently relaxed. The wavefunctions were expanded in a plane wave basis with a cutoff energy of 500 eV and a Monkhorst-Pack grid of $8\times8\times8$ k-points was used for integrations in reciprocal space for this stoichiometric conventional unit cell.
The relaxed lattice parameters of the stoichiometric conventional unit cell were then used for all other structures. For the supercell spin-polarized calculations, the force criterion was that the magnitude be less than 0.01~eV$\cdot$\AA$^{-1}$~on each atom with an energy cutoff of 500 eV.
We investigated supercells that were $2\times2\times2$, $3\times3\times3$, and $4\times4\times4$ multiples of the conventional unit cell with appropriately scaled k-point grids. 

For calculations of hyperfine constants, all carbon atoms in the supercell were taken to be $^{13}$C. We have verified that performing the calculation with $^{13}$C or $^{12}$C does not affect the converged spin density. Thus, performing the calculation with all carbon atoms as $^{13}$C will produce results for hyperfine constants at the relevant carbon atoms that are equivalent to performing calculations with only 1\% of the carbon as $^{13}$C, as dictated by their natural abundance. This result can be rationalized by the fact that hyperfine constants on the order of hundreds of MHz correspond to energies on the order of 10$^{-6}$~eV and therefore negligibly affect the convergence of electronic spin densities.   


\section{\label{discussion} RESULTS AND DISCUSSION}

For the structural features of diamond, we found a lattice constant for the conventional unit cell of $a=3.539$ \AA \ which is in good agreement with a previous theoretical calculation of $a=3.545$~\AA.~\cite{Deak} We also computed an energy  band gap of 5.40~eV that is also in good agreement with previous experimental~\cite{madelung1991semiconductors} and theoretical~\cite{Szasz} results. The general structure of X$V^-$ is shown in Fig. \ref{fig:structure} along with a numeration scheme for the identification of the nearest-neighbor carbon atoms. Bond lengths for the X$V^-$ color centers are also provided in Fig. \ref{fig:structure}.

To determine the appropriate supercell required to obtain well-converged results, we computed the hyperfine constants for one of the nearest-neighbor carbon nuclear spins ($^{13}$C$_1$) surrounding the Si$V^0$ color center as a function of supercell size. We provide these results along with $1/N$, where N is the number of atoms in a particular supercell, and $1/L$, where L is the linear supercell dimension (\textit {Cf.} Table \ref{tab:hyperfineconv}) with $L = V^{1/3}$ and $V$ is the volume of the supercell. For the $^{13}$C nuclear spins that are nearest-neighbor to the color center, the dominant contribution to the hyperfine constants will be the Fermi contact term. As shown in Eq. (\ref{eq:Atensor}), this contribution is computed by integrating over the Thomson radius, which is independent of the size of the supercell. Thus, as the bulk of the supercell is not considered in the calculation, it is more appropriate to show convergence using the scaling for the convergence of the spin density rather than using the inverse number of atoms in or the inverse volume of the supercell, since by construction the Fermi contact term cannot converge faster than the spin density. Lany and Zunger have proposed a scheme for correcting the energy of a charged defect in a supercell, which shows that the convergence of the spin density will scale as $1/L$.~\cite{Lany} Furthermore, De\'ak \textit{et al.}~\cite{Deak} have found that the Lany and Zunger scheme is appropriate for point defects in diamond. Therefore, we have considered the inverse linear supercell dimension in assessing the level of convergence of the results (including the inverse number of atoms for comparison) and find that for a 512-atom supercell the results are well-converged. 

\begin{figure}[ht!] 
\centering
\includegraphics[width=0.9\textwidth]{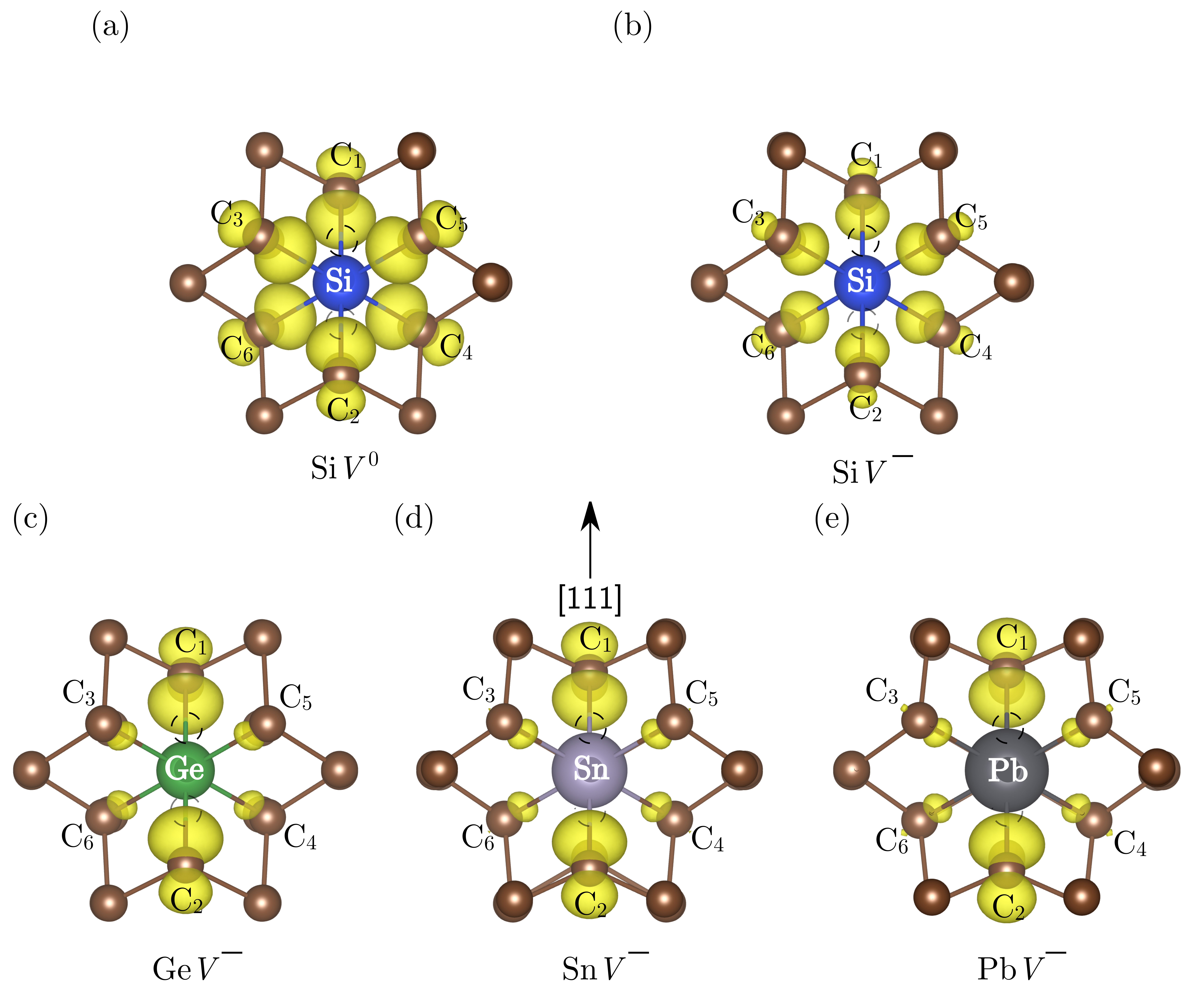}
\caption{Spin densities for (a) for Si$V^0$, (b) Si$V^-$, (c) Ge$V^-$, (d) Sn$V^-$, and (e) Pb$V^-$ with the $[111]$ direction indicated. The electron spin density is shown in yellow at an isovalue level of 0.01~$e/$\AA$^{3}$ and the carbon vacancies are shown as dashed circles.} 
\label{fig:SiV_spin_density}
\end{figure}

 Our test studies used the HSE06 functional and included the contribution of the core polarization $A_{1c}$ which is the one-center core contribution to the Fermi contact term.~\cite{gali2013ab,Szasz,Yazyev} We found that a supercell containing 512 atoms was sufficient enough to produce converged results that were in very good agreement with previous  
experimental~\cite{edmonds2008electron} and theoretical~\cite{gali2013ab} results as illustrated in Table \ref{tab:hyperfineSiV0}.
The hyperfine constants were calculated for all six of the nearest-neighbor $^{13}$C nuclear spins of Si$V^0$ and they are also in excellent agreement with previous theoretical~\cite{gali2013ab} and experimental results \cite{edmonds2008electron} (\textit {Cf.} Table \ref{tab:hyperfineSiV0}).  For the Si$V^0$ color center, we found that the hyperfine constants are evenly distributed among the nearest-neighbor $^{13}$C nuclear spins surrounding the Si atom. Furthermore, we see from Fig. \ref{fig:SiV_spin_density} that the spin density is also evenly distributed among these nearest-neighbor carbon atoms. The calculation of these hyperfine constants in a previous study~\cite{gali2013ab} has confirmed that the KUL1 center in diamond~\cite{edmonds2008electron} is actually Si$V^0$. A detailed knowledge of such hyperfine constants can also reveal how strongly the electronic spin density will interact with neighboring $^{13}$C nuclear spins.

\begin{table}[ht!]
\caption{ The hyperfine constants ($A_{xx}$, $A_{yy}$, and $A_{zz}$) in MHz for the nearest-neighbor $^{13}$C$_1$ nuclear spin of Si$V^0$ are tabulated as a function of $1/N$ and $1/L$ (in~\AA$^{-1}$), where $N$ is the number of atoms in the supercell ($N = 64,~216$ and $512$)  and $L$ is the linear supercell dimension. We use the numeration scheme defined in Fig. \ref{fig:structure} for the $^{13}$C$_1$ nuclear spin. Note that $A_{xx}$ and $A_{yy}$ are only approximately equal for small supercell sizes, due to finite size effects, which becomes more exact for larger supercell sizes. Our calculations utilize the HSE06 functional and include the contribution of core states ($A_{1c}$). In addition to $^{13}$C$_1$, we have discovered the same convergence behavior for all of the other five $^{13}$C nuclear spins surrounding Si$V^0$ as defined in Fig. \ref{fig:structure}.}
\centering
\vspace{1 mm}
\begin{tabular}{|c|c|c|c|c|}
\hline\hline 
$A_{xx}$ & $A_{yy}$ & $A_{zz}$ & $1/N$ & $1/L$\\
\hline
20 & 19 & 55 & 0.016 & 0.141\\
26 & 26 & 64 & 0.005 & 0.094\\
28 & 28 & 67 & 0.002 & 0.071\\
\hline
\end{tabular}
\label{tab:hyperfineconv}
\end{table}

\begin{table}[ht!]
\caption{ Hyperfine constants ($A_{xx}$, $A_{yy}$, and $A_{zz}$) in MHz evaluated at the six nearest-neighbor carbon nuclei surrounding the Si$V^{0}$ and Si$V^-$ impurity-vacancy color centers using the numeration scheme defined in Fig. \ref{fig:structure}. These results use a supercell of 512 atoms with the HSE06 (PBE) functional both with and without the contribution of core states ($A_{1c}$). Our results are compared with the theoretical results of Ref. \cite{gali2013ab} and the experimental results of Ref. \cite{edmonds2008electron}. The theoretical results of Ref. \cite{gali2013ab} were also performed with the HSE06 functional including the contribution of core states, but our calculations used a cutoff of 500~eV in the plane-wave expansion for the wavefunction for constant volume relaxations while Ref. \cite{gali2013ab} used 370 eV.}
\centering
\vspace{1 mm}
\begin{tabular}{|c|c|c|c|c|c|c|c|c|c|c|c|c|c|}
\hline\hline 
 \multirow{2}{*}{Color Center} &\multirow{2}{*}{Isotope} & \multicolumn{3}{|c|}{HSE06 (PBE) without $A_{1c}$} & \multicolumn{3}{|c|}{HSE06 (PBE) with $A_{1c}$} & \multicolumn{3}{|c|}{Other theory} & \multicolumn{3}{|c|}{Experiment} \\
\cline{3-14}
& & $A_{xx}$ & $A_{yy}$ & $A_{zz}$ & $A_{xx}$ & $A_{yy}$ & $A_{zz}$ & $A_{xx}$ & $A_{yy}$ & $A_{zz}$  & $A_{xx}$ & $A_{yy}$ & $A_{zz}$\\ 
\hline
  \multirow{2}{*}{Si$V^{0}$} & $^{13}$C$_{1}$, $^{13}$C$_{2}$, $^{13}$C$_{3}$ & 42 (32) & 42 (32) & 81 (65) & 28 (20) & 28 (20) & 67 (53) & 28$^{\rm a}$ & 28$^{\rm a}$ & 68$^{\rm a}$ & 30.2$^{\rm b}$ & 30.2$^{\rm b}$ & 66.2$^{\rm b}$ \\
   & $^{13}$C$_{4}$, $^{13}$C$_{5}$, $^{13}$C$_{6}$ & &  & &  &  &  &  &  &  & & &  \\
   \hline
 \multirow{3}{*}{Si$V^-$}  & $^{13}$C$_{1}$, $^{13}$C$_{2}$ & 49 (39) & 49 (39) & 89 (75) & 34 (26) & 34 (26) & 75 (62) & & & & & &   \\
  & $^{13}$C$_{3}$, $^{13}$C$_{4}$ &  49 (39) & 49 (39) & 90 (75) & 35 (26) & 35 (26) & 76 (62) & & & & & &   \\
 & $^{13}$C$_{5}$, $^{13}$C$_{6}$ & 49 (39) & 49 (39) & 90 (75) & 35 (26) & 34 (26) & 75 (62)  & & & & & & \\
\hline
\end{tabular}
\begin{flushleft}
a. From Ref. \cite{gali2013ab}

b. From Ref. \cite{edmonds2008electron}
\end{flushleft}
\label{tab:hyperfineSiV0}
\end{table}

The situation for the Si$V^-$ color center is very similar to the previously discussed Si$V^0$ case where the hyperfine constants are evenly distributed among the nearest-neighbor $^{13}$C nuclear spins surrounding the Si atom, as shown in Table \ref{tab:hyperfineSiV0}. The electron spin density is also uniformly distributed among the neighboring $^{13}$C nuclear spins in a similar manner seen for Si$V^0$ as shown in Fig. \ref{fig:SiV_spin_density}. We should note, however, that we discovered that the use of a more stringent force threshold of $10^{-4}$~eV/\AA~ for convergence of the spin-polarized calculation for the Si$V^-$ results in the appearance of some anisotropy in the calculated values of the hyperfine constants. Previous investigators have studied Si$V^0$ and Si$V^-$ defects modeled in a hydrogen-terminated carbon cluster C$_{128}$[Si$V$]H$_{98}$ with the Si$V$ defect at its center using the ORCA software package in the DFT formalism.~\cite{Nizovtsev2020} They found no significant distortion for the Si$V^0$ with a corresponding isotropic distribution of hyperfine constant strengths at the nearest-neighbor $^{13}$C nuclear spins and a distortion for the Si$V^-$ with a corresponding anisotropic distribution of hyperfine constant strengths at the nearest-neighbor $^{13}$C nuclear spins. For the specific system involving Si$V^-$, they computed the hyperfine constants $A_{xx} \approx A_{yy} \approx 105$~MHz and $A_{zz} \approx 175$~MHz for the more distant pair of nearest-neighbor $^{13}$C nuclear spins and $A_{xx} \approx A_{yy} \approx 21$~MHz and $A_{zz} \approx 37$~MHz for the closer nearest-neighbor $^{13}$C nuclear spins, while for the system involving the Si$V^0$ they found the hyperfine constants $A_{xx} \approx A_{yy} \approx 44.5$~MHz and $A_{zz} \approx 77.5$~MHz for the nearest-neighbor $^{13}$C nuclear spins.~\cite{Nizovtsev2020}

Given the agreement of our test studies for Si$V^0$ with both experiment and theory, we computed the hyperfine constants at the nearest-neighbor $^{13}$C nuclear spins of the X$V^-$ impurity-vacancy centers, where X = Ge, Sn, or Pb and $V$ is a carbon vacancy, this time using both the HSE06 and PBE functionals including the contribution of core polarization ($A_{1c}$).~\cite{gali2013ab,Szasz,Yazyev}  These hyperfine constants were calculated at the nearest-neighbor $^{13}$C nuclear spins and are listed in Table \ref{tab:hyperfineGeV-}. 
For the Ge$V^-$, we note a significant change in the behavior of the hyperfine constants when compared to those of Si$V^0$ and Si$V^-$. For example, as shown in Table \ref{tab:hyperfineGeV-}, the hyperfine constant strengths become anisotropically distributed, especially when evaluated at the $^{13}$C nuclear spins which are nearest-neighbors to the defect center. The electron spin density plots corroborate this anisotropy (\textit {Cf.} Fig. \ref{fig:SiV_spin_density}) where there are larger lobes of spin density near carbon nuclei $^{13}$C$_{1}$ and $^{13}$C$_{2}$ in contrast to the smaller lobes near $^{13}$C$_{3}$, $^{13}$C$_{4}$, $^{13}$C$_{5}$, and $^{13}$C$_{6}$. This trend is also seen for the  Sn$V^-$ and Pb$V^-$ color centers where the hyperfine constants at the nearest-neighbor $^{13}$C nuclear spins surrounding the defect center are shown in Table \ref{tab:hyperfineGeV-}. The anisotropic distributions of the electronic spin densities for the Sn$V^-$ and Pb$V^-$ centers are very similar to that seen for Ge$V^-$ as shown in Fig. \ref{fig:SiV_spin_density}. Our calculations for Ge$V^-$, Sn$V^-$, and Pb$V^-$ all reveal that, because of the anisotropic distribution in both the hyperfine constants and the electronic spin density, certain $^{13}$C nuclear spins, namely $^{13}$C$_{1}$ and $^{13}$C$_{2}$, should interact more strongly with the electronic spin density than other $^{13}$C nuclear spins, specifically $^{13}$C$_{3}$, $^{13}$C$_{4}$, $^{13}$C$_{5}$, and $^{13}$C$_{6}$.

\begin{table}[ht!]
\caption{ Hyperfine constants ($A_{xx}$, $A_{yy}$, and $A_{zz}$) in MHZ evaluated at the six nearest-neighbor carbon nuclei surrounding the Ge$V^-$, Sn$V^-$, and Pb$V^-$ impurity-vacancy color centers using the numeration scheme defined in Fig. \ref{fig:structure}. These results use a supercell of 512 atoms with the HSE06 (PBE) functional both with and without the contribution of core states ($A_{1c}$) and a cutoff of 500~eV in the plane-wave expansion for the wavefunction for constant volume relaxations.}
\centering
\vspace{1 mm}
\begin{tabular}{|c|c|c|c|c|c|c|c|}
\hline\hline 
 \multirow{2}{*}{Color Center} &\multirow{2}{*}{Isotope} & \multicolumn{3}{|c|}{HSE06 (PBE) without $A_{1c}$} & \multicolumn{3}{|c|}{HSE06 (PBE) with $A_{1c}$} \\
\cline{3-8}
& & $A_{xx}$ & $A_{yy}$ & $A_{zz}$ & $A_{xx}$ & $A_{yy}$ & $A_{zz}$ \\ 
\hline
  \multirow{3}{*}{Ge$V^-$} & $^{13}$C$_{1}$, $^{13}$C$_{2}$ & 114 (43) & 113 (42) & 209 (83) & 79 (28) & 78 (28) & 174 (68)     \\
  & $^{13}$C$_{3}$, $^{13}$C$_{4}$ & 22 (42) & 21 (42) & 42 (83) & 14 (28) & 14 (28) & 35 (68)  \\
  & $^{13}$C$_{5}$, $^{13}$C$_{6}$ & 22 (42) & 21 (42) & 42 (83) & 14 (28) & 14 (28) & 35 (68)  \\
  \hline
    \multirow{3}{*}{Sn$V^-$} & $^{13}$C$_{1}$, $^{13}$C$_{2}$  & 96 (34) & 96 (34) & 195 (77) & 59 (19) & 59 (19) & 158 (61)  \\
 & $^{13}$C$_{3}$, $^{13}$C$_{4}$  & 18 (35)  & 18 (35) & 41 (77) & 10 (19) & 10 (19) & 33 (62)  \\
 & $^{13}$C$_{5}$, $^{13}$C$_{6}$  & 18 (35) & 18 (35) & 41 (77) & 10 (19) & 10 (19) & 33 (62)  \\
 \hline
 \multirow{3}{*}{Pb$V^-$} & $^{13}$C$_{1}$, $^{13}$C$_{2}$  & 98 (32) & 97 (32) & 203 (77) & 58 (15) & 58 (15) & 163 (60)  \\
 & $^{13}$C$_{3}$, $^{13}$C$_{4}$  & 16 (32)  & 16 (32) & 38 (77) & 8 (15) & 8 (15) & 30 (60)  \\
 & $^{13}$C$_{5}$, $^{13}$C$_{6}$  & 16 (32) & 16 (32) & 38 (77) & 8 (15) & 8 (15) & 30 (60)  \\
\hline
\end{tabular}
\label{tab:hyperfineGeV-}
\end{table}



One could ask, what is the physical mechanism for this observed anisotropy in the hyperfine constants of Ge$V^-$, Sn$V^-$, and Pb$V^-$? In our calculations we found that the interatomic bond lengths among nearest-neighbor atoms and surrounding atoms beyond the defect were distorted from their equilibrium positions after relaxation due to the creation of the defect in the crystal (\textit {Cf.} Fig. \ref{fig:structure}). We recognize that distortions exist beyond the nearest-neighbor, but have chosen not to plot or discuss them here in our manuscript for the sake of clarity. These observed distortions can also be interpreted as a result of the Jahn-Teller effect as previously discussed.~\cite{Gali3}  We note that while Si$V^-$ does not show a significant amount of bond distortion, it does have the largest Jahn-Teller barrier in the ground state as noted in previous work.~\cite{Gali3} This is likely due to smaller gradients along the adiabatic potential energy surface for which a force threshold of 0.01 eV/\AA~is insufficient to explore. These distortions occur when defect structures are allowed to relax with spin-polarization where the degeneracy of the $^2E_g$ ground state is broken due to the $E\otimes e$ dynamic Jahn-Teller effect.~\cite{Gali3} Such a distortion elongates certain bonds when compared to others leading to anisotropy in the distribution of bond lengths. As suggested in the work of Gali \textit{et al.}~\cite{Gali4}, the hyperfine constants are sensitive to the distance between $^{13}$C nuclear spins and the defect center. Our calculations show that this sensitivity is so high that even minute distortions are sufficient to lead to hyperfine constants evaluated at nearest-neighbor $^{13}$C nuclear spins that differ from each other by an order of magnitude.  

For the heavier X atoms, a heuristic justification for the distortion of the bond lengths follows from the observation that the $D_{3d}$ symmetry of the X$V^-$ defects can be viewed as a weakly broken octahedral symmetry, in which case the orbitals involved in the bonding should be the $d_{xy}$, $d_{xz}$ and $d_{yz}$ orbitals. Each X atom has five valence electrons with the associated additional negative charge and each of the effective $d_{xy}$, $d_{xz}$ or $d_{yz}$ orbitals can hold at most two electrons. We use the term ``effective'' here to indicate that such orbitals may not exactly describe the system, but can serve as an approximation to allow us to elucidate its behavior. Two of these orbitals will hold two electrons and the remaining orbital will hold only one electron leading to an uneven distribution of charge among the bonds. We expect based on this distribution to see two shorter pairs of bonds and one longer bond pair, as confirmed by Fig.~\ref{fig:structure}. We note that the precision to which we have reported the bond lengths in Fig.~\ref{fig:structure} masks a distortion above numerical error that was also present for Pb$V^-$. The orbitals that hold two electrons will not contribute to the spin density, while the orbital with one electron will, thus we expect localization of the spin density on the longer bond as confirmed in Fig. \ref{fig:SiV_spin_density}.  The sensitivity of the hyperfine constants to the distance between the X atom and the carbon atom at which they are evaluated has been shown in Table \ref{tab:hyperfineGeV-} for the Ge$V^-$, Sn$V^-$, and Pb$V^-$ color centers. A similar difference is only observed for the Si$V^-$ color center when the force threshold is such that a distortion begins to emerge, leading to the notion that the difference is due to a Jahn-Teller distortion. We note here that a Jahn-Teller distortion has also been calculated for the neutral group-IV color centers~\cite{ciccarino2020strong}, but our calculations have not verified this.



Finally, we would like to comment on the anisotropy observed for our calculated hyperfine constants and their implications. First, the strengths of the hyperfine constants are anisotropically distributed among the neighboring $^{13}$C nuclear spins as the hyperfine interaction depends upon the spin density which is clearly anisotropic as shown in Fig. \ref{fig:SiV_spin_density}, due to a Jahn-Teller distortion. Next, we note that for any particular X$V^-$ center in a diamond host, its electron spin is entangled with all of the $^{13}$C nuclear spins that are present throughout the host material.~\cite{Childress281} Furthermore, it is important to make a distinction between the various types of $^{13}$C nuclear spins that surround a particular X$V^-$ center. It has been suggested that there is a barrier which we will describe by some distance r$_o$ where for $^{13}$C nuclear spins located at distances r\, $\,>$\, r$_o$, spin decoherence of the X$V^-$ center is produced because of entanglement, while for $^{13}$C nuclear spins located at distances r\, $\,<$\, r$_o$ spin coherence of the X$V^-$ center is actually enhanced.~\cite{chen2020decoherence} These two regions are separated by a barrier to spin diffusion as originally suggested by Bloembergen.~\cite{BLOEMBERGEN1949on} One interesting application of this spin diffusion barrier might be considered when we try to create a network of X$V^-$ centers in diamond experimentally. If one wishes to construct, as an illustration, a one-dimensional chain of X$V^-$centers with minimal spin decoherence among them in a network, then it is best to experimentally construct them so that the separation distance between a cluster of nearby color centers in diamond is no more than the parameter r$_o$, which describes the diffusion barrier for the particular system, and the distance to all other color centers is much greater than r$_o$. In this manner, color centers would mutually entangle and lock on to the longest spin coherence time of color centers in the cluster. For example, hybrid structures of N$V^-$ and X$V^-$ could lead to enhanced spin coherence of the X$V^-$ if the X$V^-$ is entangled with the N$V^-$. In addition, our calculations show that this parameter r$_o$ will be anisotropic when measured from an X$V^-$ center experiencing a Jahn-Teller distortion. Of course, in this discussion we make no effort to actually compute what the parameter r$_o$ is for a particular X$V^-$ color center, but simply state that it exists and that it and its angular dependence should be considered if one wishes to mitigate spin decoherence when fabricating a particular system in the laboratory.

This discussion supplements the argument that enhanced spin coherence could be obtained by lowering the temperature to suppress the fast scattering of electrons between sublevels of the ground state, mediated by phonons.~\cite{Gali3} In addition to suppressing the phonon density of states by lowering the temperature, our earlier work argues that decreasing the concentration of defects in the material can also suppress the phonon density of states.~\cite{Kuate8}

In summary, our computational results show that by using HSE06 for the exchange-correlation functional in DFT along with the inclusion of core polarization, the computed values of the hyperfine constants for the case of the Ge$V^-$, Sn$V^-$, and Pb$V^-$ color centers should be in much better agreement with experiment than by solely using the PBE exchange-correlation functional. We predict that the net effect of the increase in the hyperfine constants due to the increased localization of the defect wavefunctions from the use of the HSE06 functional and the decrease in the hyperfine constants due to the inclusion of core electron spin polarization will yield computational results that will be in better agreement with experiment as previously shown for the cases of Si$V^0$, Si$V^-$, and N$V^-$.

\section{\label{diffusivity}CONCLUSIONS}
Group-IV impurity-vacancy color centers in diamond (X$V$,  where X = Si, Ge, Sn, and Pb and $V$ is a carbon vacancy) are surrounded by $^{13}$C nuclear spins in diamond which possess a nonzero spin which can limit the spin coherence properties of these X$V$ color centers. We have explored the dependence of the hyperfine constants on the location of $^{13}$C nuclear spins that are nearest-neighbors to group-IV color centers. As has been seen for the case of Si$V^0$, we have shown that the use of the hybrid functional HSE06 for exchange and correlation over the use of the PBE functional provides a far more accurate means of computing the hyperfine constants and we have extended these calculations to Ge$V^-$, Sn$V^-$, and Pb$V^-$. It is the hope that our results will serve as a useful benchmark in both seeing how to improve DFT studies for these impurity-vacancy systems and will provide further guidance for future experimental research on these systems.

We find that the distribution of the hyperfine constants is anisotropic when the color centers exhibit a Jahn-Teller distortion. We propose this effect can be exploited to mitigate spin decoherence of the color centers by repeating cooling procedures such that the color center is cycled between the three energetically equivalent distortions, ultimately finding one that favors enhanced spin coherence. We additionally argue that consideration of a direction-dependent spin diffusion barrier must be made in constructing hybrid systems of color centers.     

{\bf ACKNOWLEDGMENTS:}
We acknowledge support by the STC Center for Integrated Quantum Materials, NSF Grant No. DMR-1231319. R.K.D. gratefully acknowledges financial support from the Princeton Presidential Postdoctoral Research Fellowship. This work used computational resources of the Extreme Science and Engineering Discovery Environment (XSEDE), which is supported by National Science Foundation Grant Number ACI-1548562,~\cite{Towns} on Stampede2 at TACC through allocation TG-DMR120073, and of the National Energy Research Scientific Computing Center (NERSC), a U.S. Department of Energy Office of Science User Facility operated under Contract No. DE-AC02-05CH11231.  We would like to thank Marko Lon\ifmmode \check{c}\else \v{c}\fi{}ar for bringing our attention to Ref.~\cite{ciccarino2020strong}.

\bibliography{refs_ESR-XV-Diamond}

\begin{thebibliography}{65}%
\makeatletter
\providecommand \@ifxundefined [1]{%
 \@ifx{#1\undefined}
}%
\providecommand \@ifnum [1]{%
 \ifnum #1\expandafter \@firstoftwo
 \else \expandafter \@secondoftwo
 \fi
}%
\providecommand \@ifx [1]{%
 \ifx #1\expandafter \@firstoftwo
 \else \expandafter \@secondoftwo
 \fi
}%
\providecommand \natexlab [1]{#1}%
\providecommand \enquote  [1]{``#1''}%
\providecommand \bibnamefont  [1]{#1}%
\providecommand \bibfnamefont [1]{#1}%
\providecommand \citenamefont [1]{#1}%
\providecommand \href@noop [0]{\@secondoftwo}%
\providecommand \href [0]{\begingroup \@sanitize@url \@href}%
\providecommand \@href[1]{\@@startlink{#1}\@@href}%
\providecommand \@@href[1]{\endgroup#1\@@endlink}%
\providecommand \@sanitize@url [0]{\catcode `\\12\catcode `\$12\catcode
  `\&12\catcode `\#12\catcode `\^12\catcode `\_12\catcode `\%12\relax}%
\providecommand \@@startlink[1]{}%
\providecommand \@@endlink[0]{}%
\providecommand \url  [0]{\begingroup\@sanitize@url \@url }%
\providecommand \@url [1]{\endgroup\@href {#1}{\urlprefix }}%
\providecommand \urlprefix  [0]{URL }%
\providecommand \Eprint [0]{\href }%
\providecommand \doibase [0]{https://doi.org/}%
\providecommand \selectlanguage [0]{\@gobble}%
\providecommand \bibinfo  [0]{\@secondoftwo}%
\providecommand \bibfield  [0]{\@secondoftwo}%
\providecommand \translation [1]{[#1]}%
\providecommand \BibitemOpen [0]{}%
\providecommand \bibitemStop [0]{}%
\providecommand \bibitemNoStop [0]{.\EOS\space}%
\providecommand \EOS [0]{\spacefactor3000\relax}%
\providecommand \BibitemShut  [1]{\csname bibitem#1\endcsname}%
\let\auto@bib@innerbib\@empty
\bibitem [{\citenamefont {Ewen$~$}\ and\ \citenamefont
  {Purcell}(1951)}]{Purcell1951}%
  \BibitemOpen
  \bibfield  {author} {\bibinfo {author} {\bibfnamefont {H.~I.}\ \bibnamefont
  {Ewen$~$}}and\ \bibinfo {author} {\bibfnamefont {E.~M.}\ \bibnamefont
  {Purcell}},\ }\href {https://doi.org/10.1038/168356a0} {\bibfield  {journal}
  {\bibinfo  {journal} {Nature}\ }\textbf {\bibinfo {volume} {168}},\ \bibinfo
  {pages} {356} (\bibinfo {year} {1951})}\BibitemShut {NoStop}%
\bibitem [{\citenamefont {Muller$~$}\ and\ \citenamefont
  {Oort}(1951)}]{Muller1951}%
  \BibitemOpen
  \bibfield  {author} {\bibinfo {author} {\bibfnamefont {C.~A.}\ \bibnamefont
  {Muller$~$}}and\ \bibinfo {author} {\bibfnamefont {J.~H.}\ \bibnamefont
  {Oort}},\ }\href {https://doi.org/10.1038/168357a0} {\bibfield  {journal}
  {\bibinfo  {journal} {Nature}\ }\textbf {\bibinfo {volume} {168}},\ \bibinfo
  {pages} {357} (\bibinfo {year} {1951})}\BibitemShut {NoStop}%
\bibitem [{\citenamefont {Kittel}(2004)}]{kittel2004introduction}%
  \BibitemOpen
  \bibfield  {author} {\bibinfo {author} {\bibfnamefont {C.}~\bibnamefont
  {Kittel}},\ }\href {https://books.google.com/books?id=kym4QgAACAAJ} {\emph
  {\bibinfo {title} {Introduction to Solid State Physics}}}\ (\bibinfo
  {publisher} {Wiley},\ \bibinfo {year} {2004})\BibitemShut {NoStop}%
\bibitem [{\citenamefont {Dreher}\ \emph {et~al.}(2011)\citenamefont {Dreher},
  \citenamefont {Hilker}, \citenamefont {Brandlmaier}, \citenamefont
  {Goennenwein}, \citenamefont {Huebl}, \citenamefont {Stutzmann},\ and\
  \citenamefont {Brandt}}]{Dreher}%
  \BibitemOpen
  \bibfield  {author} {\bibinfo {author} {\bibfnamefont {L.}~\bibnamefont
  {Dreher}}, \bibinfo {author} {\bibfnamefont {T.~A.}\ \bibnamefont {Hilker}},
  \bibinfo {author} {\bibfnamefont {A.}~\bibnamefont {Brandlmaier}}, \bibinfo
  {author} {\bibfnamefont {S.~T.~B.}\ \bibnamefont {Goennenwein}}, \bibinfo
  {author} {\bibfnamefont {H.}~\bibnamefont {Huebl}}, \bibinfo {author}
  {\bibfnamefont {M.}~\bibnamefont {Stutzmann}}, and\ \bibinfo {author}
  {\bibfnamefont {M.~S.}\ \bibnamefont {Brandt}},\ }\href
  {https://doi.org/10.1103/PhysRevLett.106.037601} {\bibfield  {journal}
  {\bibinfo  {journal} {Phys. Rev. Lett.}\ }\textbf {\bibinfo {volume} {106}},\
  \bibinfo {pages} {037601} (\bibinfo {year} {2011})}\BibitemShut {NoStop}%
\bibitem [{\citenamefont {Griffiths}(1982)}]{griffiths1982hyperfine}%
  \BibitemOpen
  \bibfield  {author} {\bibinfo {author} {\bibfnamefont {D.~J.}\ \bibnamefont
  {Griffiths}},\ }\href {https://doi.org/10.1119/1.12733} {\bibfield  {journal}
  {\bibinfo  {journal} {American Journal of physics}\ }\textbf {\bibinfo
  {volume} {50}},\ \bibinfo {pages} {698} (\bibinfo {year} {1982})}\BibitemShut
  {NoStop}%
\bibitem [{\citenamefont {Griffiths}(2014)}]{griffiths2014introduction}%
  \BibitemOpen
  \bibfield  {author} {\bibinfo {author} {\bibfnamefont {D.}~\bibnamefont
  {Griffiths}},\ }\href {https://books.google.com/books?id=J9ygBwAAQBAJ} {\emph
  {\bibinfo {title} {Introduction to Electrodynamics}}}\ (\bibinfo  {publisher}
  {Pearson Education},\ \bibinfo {year} {2014})\BibitemShut {NoStop}%
\bibitem [{\citenamefont {Slichter}(2013)}]{slichter2013principles}%
  \BibitemOpen
  \bibfield  {author} {\bibinfo {author} {\bibfnamefont {C.~P.}\ \bibnamefont
  {Slichter}},\ }\href@noop {} {\emph {\bibinfo {title} {Principles of magnetic
  resonance}}},\ Vol.~\bibinfo {volume} {1}\ (\bibinfo  {publisher} {Springer
  Science \& Business Media},\ \bibinfo {year} {2013})\BibitemShut {NoStop}%
\bibitem [{\citenamefont {Bl\"ochl}(2000)}]{Blochl2000}%
  \BibitemOpen
  \bibfield  {author} {\bibinfo {author} {\bibfnamefont {P.~E.}\ \bibnamefont
  {Bl\"ochl}},\ }\href {https://doi.org/10.1103/PhysRevB.62.6158} {\bibfield
  {journal} {\bibinfo  {journal} {Phys. Rev. B}\ }\textbf {\bibinfo {volume}
  {62}},\ \bibinfo {pages} {6158} (\bibinfo {year} {2000})}\BibitemShut
  {NoStop}%
\bibitem [{\citenamefont {Gali}\ \emph {et~al.}(2008)\citenamefont {Gali},
  \citenamefont {Fyta},\ and\ \citenamefont {Kaxiras}}]{Gali4}%
  \BibitemOpen
  \bibfield  {author} {\bibinfo {author} {\bibfnamefont {A.}~\bibnamefont
  {Gali}}, \bibinfo {author} {\bibfnamefont {M.}~\bibnamefont {Fyta}}, and\
  \bibinfo {author} {\bibfnamefont {E.}~\bibnamefont {Kaxiras}},\ }\href
  {https://doi.org/10.1103/PhysRevB.77.155206} {\bibfield  {journal} {\bibinfo
  {journal} {Phys. Rev. B}\ }\textbf {\bibinfo {volume} {77}},\ \bibinfo
  {pages} {155206} (\bibinfo {year} {2008})}\BibitemShut {NoStop}%
\bibitem [{\citenamefont {P\'erez~Galv\'an}\ \emph {et~al.}(2008)\citenamefont
  {P\'erez~Galv\'an}, \citenamefont {Zhao},\ and\ \citenamefont
  {Orozco}}]{Galvan}%
  \BibitemOpen
  \bibfield  {author} {\bibinfo {author} {\bibfnamefont {A.}~\bibnamefont
  {P\'erez~Galv\'an}}, \bibinfo {author} {\bibfnamefont {Y.}~\bibnamefont
  {Zhao}}, and\ \bibinfo {author} {\bibfnamefont {L.~A.}\ \bibnamefont
  {Orozco}},\ }\href {https://doi.org/10.1103/PhysRevA.78.012502} {\bibfield
  {journal} {\bibinfo  {journal} {Phys. Rev. A}\ }\textbf {\bibinfo {volume}
  {78}},\ \bibinfo {pages} {012502} (\bibinfo {year} {2008})}\BibitemShut
  {NoStop}%
\bibitem [{\citenamefont {Van~de Walle$~$}\ and\ \citenamefont
  {Bl{\"o}chl}(1993)}]{Walle}%
  \BibitemOpen
  \bibfield  {author} {\bibinfo {author} {\bibfnamefont {C.~G.}\ \bibnamefont
  {Van~de Walle$~$}}and\ \bibinfo {author} {\bibfnamefont {P.~E.}\ \bibnamefont
  {Bl{\"o}chl}},\ }\href {https://doi.org/10.1103/PhysRevB.47.4244} {\bibfield
  {journal} {\bibinfo  {journal} {Phys. Rev. B}\ }\textbf {\bibinfo {volume}
  {47}},\ \bibinfo {pages} {4244} (\bibinfo {year} {1993})}\BibitemShut
  {NoStop}%
\bibitem [{\citenamefont {Sz\'asz}\ \emph {et~al.}(2013)\citenamefont
  {Sz\'asz}, \citenamefont {Hornos}, \citenamefont {Marsman},\ and\
  \citenamefont {Gali}}]{Szasz}%
  \BibitemOpen
  \bibfield  {author} {\bibinfo {author} {\bibfnamefont {K.}~\bibnamefont
  {Sz\'asz}}, \bibinfo {author} {\bibfnamefont {T.}~\bibnamefont {Hornos}},
  \bibinfo {author} {\bibfnamefont {M.}~\bibnamefont {Marsman}}, and\ \bibinfo
  {author} {\bibfnamefont {A.}~\bibnamefont {Gali}},\ }\href
  {https://doi.org/10.1103/PhysRevB.88.075202} {\bibfield  {journal} {\bibinfo
  {journal} {Phys. Rev. B}\ }\textbf {\bibinfo {volume} {88}},\ \bibinfo
  {pages} {075202} (\bibinfo {year} {2013})}\BibitemShut {NoStop}%
\bibitem [{\citenamefont {Childress}\ \emph {et~al.}(2006)\citenamefont
  {Childress}, \citenamefont {Gurudev~Dutt}, \citenamefont {Taylor},
  \citenamefont {Zibrov}, \citenamefont {Jelezko}, \citenamefont {Wrachtrup},
  \citenamefont {Hemmer},\ and\ \citenamefont {Lukin}}]{Childress281}%
  \BibitemOpen
  \bibfield  {author} {\bibinfo {author} {\bibfnamefont {L.}~\bibnamefont
  {Childress}}, \bibinfo {author} {\bibfnamefont {M.~V.}\ \bibnamefont
  {Gurudev~Dutt}}, \bibinfo {author} {\bibfnamefont {J.~M.}\ \bibnamefont
  {Taylor}}, \bibinfo {author} {\bibfnamefont {A.~S.}\ \bibnamefont {Zibrov}},
  \bibinfo {author} {\bibfnamefont {F.}~\bibnamefont {Jelezko}}, \bibinfo
  {author} {\bibfnamefont {J.}~\bibnamefont {Wrachtrup}}, \bibinfo {author}
  {\bibfnamefont {P.~R.}\ \bibnamefont {Hemmer}}, and\ \bibinfo {author}
  {\bibfnamefont {M.~D.}\ \bibnamefont {Lukin}},\ }\href
  {https://doi.org/10.1126/science.1131871} {\bibfield  {journal} {\bibinfo
  {journal} {Science}\ }\textbf {\bibinfo {volume} {314}},\ \bibinfo {pages}
  {281} (\bibinfo {year} {2006})}\BibitemShut {NoStop}%
\bibitem [{\citenamefont {Doherty}\ \emph {et~al.}(2013)\citenamefont
  {Doherty}, \citenamefont {Manson}, \citenamefont {Delaney}, \citenamefont
  {Jelezko}, \citenamefont {Wrachtrup},\ and\ \citenamefont
  {Hollenberg}}]{Doherty2013nitrogen}%
  \BibitemOpen
  \bibfield  {author} {\bibinfo {author} {\bibfnamefont {M.~W.}\ \bibnamefont
  {Doherty}}, \bibinfo {author} {\bibfnamefont {N.~B.}\ \bibnamefont {Manson}},
  \bibinfo {author} {\bibfnamefont {P.}~\bibnamefont {Delaney}}, \bibinfo
  {author} {\bibfnamefont {F.}~\bibnamefont {Jelezko}}, \bibinfo {author}
  {\bibfnamefont {J.}~\bibnamefont {Wrachtrup}}, and\ \bibinfo {author}
  {\bibfnamefont {L.~C.}\ \bibnamefont {Hollenberg}},\ }\href
  {https://doi.org/https://doi.org/10.1016/j.physrep.2013.02.001} {\bibfield
  {journal} {\bibinfo  {journal} {Phys. Rep.}\ }\textbf {\bibinfo {volume}
  {528}},\ \bibinfo {pages} {1} (\bibinfo {year} {2013})}\BibitemShut {NoStop}%
\bibitem [{\citenamefont {Awschalom}\ \emph {et~al.}(2018)\citenamefont
  {Awschalom}, \citenamefont {Hanson}, \citenamefont {Wrachtrup},\ and\
  \citenamefont {Zhou}}]{awschalom2018quant}%
  \BibitemOpen
  \bibfield  {author} {\bibinfo {author} {\bibfnamefont {D.~D.}\ \bibnamefont
  {Awschalom}}, \bibinfo {author} {\bibfnamefont {R.}~\bibnamefont {Hanson}},
  \bibinfo {author} {\bibfnamefont {J.}~\bibnamefont {Wrachtrup}}, and\
  \bibinfo {author} {\bibfnamefont {B.~B.}\ \bibnamefont {Zhou}},\ }\href
  {https://doi.org/10.1038/s41566-018-0232-2} {\bibfield  {journal} {\bibinfo
  {journal} {Nat. Photon.}\ }\textbf {\bibinfo {volume} {12}},\ \bibinfo
  {pages} {516} (\bibinfo {year} {2018})}\BibitemShut {NoStop}%
\bibitem [{\citenamefont {Kuate~Defo}\ \emph
  {et~al.}(2019{\natexlab{a}})\citenamefont {Kuate~Defo}, \citenamefont
  {Wang},\ and\ \citenamefont {Manjunathaiah}}]{DEFO2019}%
  \BibitemOpen
  \bibfield  {author} {\bibinfo {author} {\bibfnamefont {R.}~\bibnamefont
  {Kuate~Defo}}, \bibinfo {author} {\bibfnamefont {R.}~\bibnamefont {Wang}},
  and\ \bibinfo {author} {\bibfnamefont {M.}~\bibnamefont {Manjunathaiah}},\
  }\href {http://www.sciencedirect.com/science/article/pii/S1877750318304320}
  {\bibfield  {journal} {\bibinfo  {journal} {Journal of Computational
  Science}\ } (\bibinfo {year} {2019}{\natexlab{a}})}\BibitemShut {NoStop}%
\bibitem [{\citenamefont {Kuate~Defo}\ \emph {et~al.}(2018)\citenamefont
  {Kuate~Defo}, \citenamefont {Zhang}, \citenamefont {Bracher}, \citenamefont
  {Kim}, \citenamefont {Hu},\ and\ \citenamefont {Kaxiras}}]{Kuate}%
  \BibitemOpen
  \bibfield  {author} {\bibinfo {author} {\bibfnamefont {R.}~\bibnamefont
  {Kuate~Defo}}, \bibinfo {author} {\bibfnamefont {X.}~\bibnamefont {Zhang}},
  \bibinfo {author} {\bibfnamefont {D.}~\bibnamefont {Bracher}}, \bibinfo
  {author} {\bibfnamefont {G.}~\bibnamefont {Kim}}, \bibinfo {author}
  {\bibfnamefont {E.}~\bibnamefont {Hu}}, and\ \bibinfo {author} {\bibfnamefont
  {E.}~\bibnamefont {Kaxiras}},\ }\href
  {https://doi.org/10.1103/PhysRevB.98.104103} {\bibfield  {journal} {\bibinfo
  {journal} {Phys. Rev. B}\ }\textbf {\bibinfo {volume} {98}},\ \bibinfo
  {pages} {104103} (\bibinfo {year} {2018})}\BibitemShut {NoStop}%
\bibitem [{\citenamefont {Kuate~Defo}\ \emph
  {et~al.}(2019{\natexlab{b}})\citenamefont {Kuate~Defo}, \citenamefont
  {Kaxiras},\ and\ \citenamefont {Richardson}}]{Kuate8}%
  \BibitemOpen
  \bibfield  {author} {\bibinfo {author} {\bibfnamefont {R.}~\bibnamefont
  {Kuate~Defo}}, \bibinfo {author} {\bibfnamefont {E.}~\bibnamefont {Kaxiras}},
  and\ \bibinfo {author} {\bibfnamefont {S.~L.}\ \bibnamefont {Richardson}},\
  }\href {https://doi.org/10.1063/1.5123227} {\bibfield  {journal} {\bibinfo
  {journal} {Journal of Applied Physics}\ }\textbf {\bibinfo {volume} {126}},\
  \bibinfo {pages} {195103} (\bibinfo {year} {2019}{\natexlab{b}})}\BibitemShut
  {NoStop}%
\bibitem [{\citenamefont {Ladd}\ \emph {et~al.}(2010)\citenamefont {Ladd},
  \citenamefont {Jelezko}, \citenamefont {Laflamme}, \citenamefont {Nakamura},
  \citenamefont {Monroe},\ and\ \citenamefont {O'Brien}}]{Ladd2010}%
  \BibitemOpen
  \bibfield  {author} {\bibinfo {author} {\bibfnamefont {T.~D.}\ \bibnamefont
  {Ladd}}, \bibinfo {author} {\bibfnamefont {F.}~\bibnamefont {Jelezko}},
  \bibinfo {author} {\bibfnamefont {R.}~\bibnamefont {Laflamme}}, \bibinfo
  {author} {\bibfnamefont {Y.}~\bibnamefont {Nakamura}}, \bibinfo {author}
  {\bibfnamefont {C.}~\bibnamefont {Monroe}}, and\ \bibinfo {author}
  {\bibfnamefont {J.~L.}\ \bibnamefont {O'Brien}},\ }\href
  {https://doi.org/10.1038/nature08812} {\bibfield  {journal} {\bibinfo
  {journal} {Nature}\ }\textbf {\bibinfo {volume} {464}},\ \bibinfo {pages}
  {45} (\bibinfo {year} {2010})}\BibitemShut {NoStop}%
\bibitem [{\citenamefont {Pfaff}\ \emph {et~al.}(2014)\citenamefont {Pfaff},
  \citenamefont {Hensen}, \citenamefont {Bernien}, \citenamefont {van Dam},
  \citenamefont {Blok}, \citenamefont {Taminiau}, \citenamefont {Tiggelman},
  \citenamefont {Schouten}, \citenamefont {Markham}, \citenamefont {Twitchen},\
  and\ \citenamefont {Hanson}}]{Pfaff532}%
  \BibitemOpen
  \bibfield  {author} {\bibinfo {author} {\bibfnamefont {W.}~\bibnamefont
  {Pfaff}}, \bibinfo {author} {\bibfnamefont {B.~J.}\ \bibnamefont {Hensen}},
  \bibinfo {author} {\bibfnamefont {H.}~\bibnamefont {Bernien}}, \bibinfo
  {author} {\bibfnamefont {S.~B.}\ \bibnamefont {van Dam}}, \bibinfo {author}
  {\bibfnamefont {M.~S.}\ \bibnamefont {Blok}}, \bibinfo {author}
  {\bibfnamefont {T.~H.}\ \bibnamefont {Taminiau}}, \bibinfo {author}
  {\bibfnamefont {M.~J.}\ \bibnamefont {Tiggelman}}, \bibinfo {author}
  {\bibfnamefont {R.~N.}\ \bibnamefont {Schouten}}, \bibinfo {author}
  {\bibfnamefont {M.}~\bibnamefont {Markham}}, \bibinfo {author} {\bibfnamefont
  {D.~J.}\ \bibnamefont {Twitchen}}, and\ \bibinfo {author} {\bibfnamefont
  {R.}~\bibnamefont {Hanson}},\ }\href
  {https://doi.org/10.1126/science.1253512} {\bibfield  {journal} {\bibinfo
  {journal} {Science}\ }\textbf {\bibinfo {volume} {345}},\ \bibinfo {pages}
  {532} (\bibinfo {year} {2014})}\BibitemShut {NoStop}%
\bibitem [{\citenamefont {Ceperley$~$}\ and\ \citenamefont
  {Alder}(1980)}]{ceperley1980ground}%
  \BibitemOpen
  \bibfield  {author} {\bibinfo {author} {\bibfnamefont {D.~M.}\ \bibnamefont
  {Ceperley$~$}}and\ \bibinfo {author} {\bibfnamefont {B.~J.}\ \bibnamefont
  {Alder}},\ }\href {https://doi.org/10.1103/PhysRevLett.45.566} {\bibfield
  {journal} {\bibinfo  {journal} {Phys. Rev. Lett.}\ }\textbf {\bibinfo
  {volume} {45}},\ \bibinfo {pages} {566} (\bibinfo {year} {1980})}\BibitemShut
  {NoStop}%
\bibitem [{\citenamefont {Perdew$~$}\ and\ \citenamefont
  {Zunger}(1981)}]{perdew1981self}%
  \BibitemOpen
  \bibfield  {author} {\bibinfo {author} {\bibfnamefont {J.~P.}\ \bibnamefont
  {Perdew$~$}}and\ \bibinfo {author} {\bibfnamefont {A.}~\bibnamefont
  {Zunger}},\ }\href {https://doi.org/10.1103/PhysRevB.23.5048} {\bibfield
  {journal} {\bibinfo  {journal} {Phys. Rev. B}\ }\textbf {\bibinfo {volume}
  {23}},\ \bibinfo {pages} {5048} (\bibinfo {year} {1981})}\BibitemShut
  {NoStop}%
\bibitem [{\citenamefont {Gali}(2009)}]{GaliPBE}%
  \BibitemOpen
  \bibfield  {author} {\bibinfo {author} {\bibfnamefont {A.}~\bibnamefont
  {Gali}},\ }\href {https://doi.org/10.1103/PhysRevB.80.241204} {\bibfield
  {journal} {\bibinfo  {journal} {Phys. Rev. B}\ }\textbf {\bibinfo {volume}
  {80}},\ \bibinfo {pages} {241204} (\bibinfo {year} {2009})}\BibitemShut
  {NoStop}%
\bibitem [{\citenamefont {Perdew}\ \emph {et~al.}(1996)\citenamefont {Perdew},
  \citenamefont {Burke},\ and\ \citenamefont {Ernzerhof}}]{Perdew}%
  \BibitemOpen
  \bibfield  {author} {\bibinfo {author} {\bibfnamefont {J.~P.}\ \bibnamefont
  {Perdew}}, \bibinfo {author} {\bibfnamefont {K.}~\bibnamefont {Burke}}, and\
  \bibinfo {author} {\bibfnamefont {M.}~\bibnamefont {Ernzerhof}},\ }\href
  {https://doi.org/10.1103/PhysRevLett.77.3865} {\bibfield  {journal} {\bibinfo
   {journal} {Phys. Rev. Lett.}\ }\textbf {\bibinfo {volume} {77}},\ \bibinfo
  {pages} {3865} (\bibinfo {year} {1996})}\BibitemShut {NoStop}%
\bibitem [{\citenamefont {Smeltzer}\ \emph {et~al.}(2011)\citenamefont
  {Smeltzer}, \citenamefont {Childress},\ and\ \citenamefont
  {Gali}}]{Smeltzer_2011}%
  \BibitemOpen
  \bibfield  {author} {\bibinfo {author} {\bibfnamefont {B.}~\bibnamefont
  {Smeltzer}}, \bibinfo {author} {\bibfnamefont {L.}~\bibnamefont {Childress}},
  and\ \bibinfo {author} {\bibfnamefont {A.}~\bibnamefont {Gali}},\ }\href
  {https://doi.org/10.1088/1367-2630/13/2/025021} {\bibfield  {journal}
  {\bibinfo  {journal} {New Journal of Physics}\ }\textbf {\bibinfo {volume}
  {13}},\ \bibinfo {pages} {025021} (\bibinfo {year} {2011})}\BibitemShut
  {NoStop}%
\bibitem [{\citenamefont {Gali}\ \emph {et~al.}(2009)\citenamefont {Gali},
  \citenamefont {Janz\'en}, \citenamefont {De\'ak}, \citenamefont {Kresse},\
  and\ \citenamefont {Kaxiras}}]{Gali2}%
  \BibitemOpen
  \bibfield  {author} {\bibinfo {author} {\bibfnamefont {A.}~\bibnamefont
  {Gali}}, \bibinfo {author} {\bibfnamefont {E.}~\bibnamefont {Janz\'en}},
  \bibinfo {author} {\bibfnamefont {P.}~\bibnamefont {De\'ak}}, \bibinfo
  {author} {\bibfnamefont {G.}~\bibnamefont {Kresse}}, and\ \bibinfo {author}
  {\bibfnamefont {E.}~\bibnamefont {Kaxiras}},\ }\href
  {https://doi.org/10.1103/PhysRevLett.103.186404} {\bibfield  {journal}
  {\bibinfo  {journal} {Phys. Rev. Lett.}\ }\textbf {\bibinfo {volume} {103}},\
  \bibinfo {pages} {186404} (\bibinfo {year} {2009})}\BibitemShut {NoStop}%
\bibitem [{\citenamefont {Yazyev}\ \emph {et~al.}(2005)\citenamefont {Yazyev},
  \citenamefont {Tavernelli}, \citenamefont {Helm},\ and\ \citenamefont
  {R\"othlisberger}}]{Yazyev}%
  \BibitemOpen
  \bibfield  {author} {\bibinfo {author} {\bibfnamefont {O.~V.}\ \bibnamefont
  {Yazyev}}, \bibinfo {author} {\bibfnamefont {I.}~\bibnamefont {Tavernelli}},
  \bibinfo {author} {\bibfnamefont {L.}~\bibnamefont {Helm}}, and\ \bibinfo
  {author} {\bibfnamefont {U.}~\bibnamefont {R\"othlisberger}},\ }\href
  {https://doi.org/10.1103/PhysRevB.71.115110} {\bibfield  {journal} {\bibinfo
  {journal} {Phys. Rev. B}\ }\textbf {\bibinfo {volume} {71}},\ \bibinfo
  {pages} {115110} (\bibinfo {year} {2005})}\BibitemShut {NoStop}%
\bibitem [{\citenamefont {Kresse$~$}\ and\ \citenamefont
  {Hafner}(1993)}]{Kresse1}%
  \BibitemOpen
  \bibfield  {author} {\bibinfo {author} {\bibfnamefont {G.}~\bibnamefont
  {Kresse$~$}}and\ \bibinfo {author} {\bibfnamefont {J.}~\bibnamefont
  {Hafner}},\ }\href {https://doi.org/10.1103/PhysRevB.47.558} {\bibfield
  {journal} {\bibinfo  {journal} {Phys. Rev. B}\ }\textbf {\bibinfo {volume}
  {47}},\ \bibinfo {pages} {558} (\bibinfo {year} {1993})}\BibitemShut
  {NoStop}%
\bibitem [{\citenamefont {Kresse$~$}\ and\ \citenamefont
  {Furthm\"uller}(1996)}]{Kresse2}%
  \BibitemOpen
  \bibfield  {author} {\bibinfo {author} {\bibfnamefont {G.}~\bibnamefont
  {Kresse$~$}}and\ \bibinfo {author} {\bibfnamefont {J.}~\bibnamefont
  {Furthm\"uller}},\ }\href {https://doi.org/10.1103/PhysRevB.54.11169}
  {\bibfield  {journal} {\bibinfo  {journal} {Phys. Rev. B}\ }\textbf {\bibinfo
  {volume} {54}},\ \bibinfo {pages} {11169} (\bibinfo {year}
  {1996})}\BibitemShut {NoStop}%
\bibitem [{\citenamefont {Kresse$~$}\ and\ \citenamefont
  {Joubert}(1999)}]{Kresse3}%
  \BibitemOpen
  \bibfield  {author} {\bibinfo {author} {\bibfnamefont {G.}~\bibnamefont
  {Kresse$~$}}and\ \bibinfo {author} {\bibfnamefont {D.}~\bibnamefont
  {Joubert}},\ }\href {https://doi.org/10.1103/PhysRevB.59.1758} {\bibfield
  {journal} {\bibinfo  {journal} {Phys. Rev. B}\ }\textbf {\bibinfo {volume}
  {59}},\ \bibinfo {pages} {1758} (\bibinfo {year} {1999})}\BibitemShut
  {NoStop}%
\bibitem [{\citenamefont {Heyd}\ \emph {et~al.}(2003)\citenamefont {Heyd},
  \citenamefont {Scuseria},\ and\ \citenamefont {Ernzerhof}}]{Heyd}%
  \BibitemOpen
  \bibfield  {author} {\bibinfo {author} {\bibfnamefont {J.}~\bibnamefont
  {Heyd}}, \bibinfo {author} {\bibfnamefont {G.~E.}\ \bibnamefont {Scuseria}},
  and\ \bibinfo {author} {\bibfnamefont {M.}~\bibnamefont {Ernzerhof}},\ }\href
  {https://doi.org/10.1063/1.1564060} {\bibfield  {journal} {\bibinfo
  {journal} {The Journal of Chemical Physics}\ }\textbf {\bibinfo {volume}
  {118}},\ \bibinfo {pages} {8207} (\bibinfo {year} {2003})}\BibitemShut
  {NoStop}%
\bibitem [{\citenamefont {Krukau}\ \emph {et~al.}(2006)\citenamefont {Krukau},
  \citenamefont {Vydrov}, \citenamefont {Izmaylov},\ and\ \citenamefont
  {Scuseria}}]{Krukau}%
  \BibitemOpen
  \bibfield  {author} {\bibinfo {author} {\bibfnamefont {A.~V.}\ \bibnamefont
  {Krukau}}, \bibinfo {author} {\bibfnamefont {O.~A.}\ \bibnamefont {Vydrov}},
  \bibinfo {author} {\bibfnamefont {A.~F.}\ \bibnamefont {Izmaylov}}, and\
  \bibinfo {author} {\bibfnamefont {G.~E.}\ \bibnamefont {Scuseria}},\ }\href
  {https://doi.org/10.1063/1.2404663} {\bibfield  {journal} {\bibinfo
  {journal} {The Journal of Chemical Physics}\ }\textbf {\bibinfo {volume}
  {125}},\ \bibinfo {pages} {224106} (\bibinfo {year} {2006})}\BibitemShut
  {NoStop}%
\bibitem [{\citenamefont {Felton}\ \emph {et~al.}(2009)\citenamefont {Felton},
  \citenamefont {Edmonds}, \citenamefont {Newton}, \citenamefont {Martineau},
  \citenamefont {Fisher}, \citenamefont {Twitchen},\ and\ \citenamefont
  {Baker}}]{Felton}%
  \BibitemOpen
  \bibfield  {author} {\bibinfo {author} {\bibfnamefont {S.}~\bibnamefont
  {Felton}}, \bibinfo {author} {\bibfnamefont {A.~M.}\ \bibnamefont {Edmonds}},
  \bibinfo {author} {\bibfnamefont {M.~E.}\ \bibnamefont {Newton}}, \bibinfo
  {author} {\bibfnamefont {P.~M.}\ \bibnamefont {Martineau}}, \bibinfo {author}
  {\bibfnamefont {D.}~\bibnamefont {Fisher}}, \bibinfo {author} {\bibfnamefont
  {D.~J.}\ \bibnamefont {Twitchen}}, and\ \bibinfo {author} {\bibfnamefont
  {J.~M.}\ \bibnamefont {Baker}},\ }\href
  {https://doi.org/10.1103/PhysRevB.79.075203} {\bibfield  {journal} {\bibinfo
  {journal} {Phys. Rev. B}\ }\textbf {\bibinfo {volume} {79}},\ \bibinfo
  {pages} {075203} (\bibinfo {year} {2009})}\BibitemShut {NoStop}%
\bibitem [{\citenamefont {De\'ak}\ \emph {et~al.}(2014)\citenamefont {De\'ak},
  \citenamefont {Aradi}, \citenamefont {Kaviani}, \citenamefont {Frauenheim},\
  and\ \citenamefont {Gali}}]{Deak}%
  \BibitemOpen
  \bibfield  {author} {\bibinfo {author} {\bibfnamefont {P.}~\bibnamefont
  {De\'ak}}, \bibinfo {author} {\bibfnamefont {B.}~\bibnamefont {Aradi}},
  \bibinfo {author} {\bibfnamefont {M.}~\bibnamefont {Kaviani}}, \bibinfo
  {author} {\bibfnamefont {T.}~\bibnamefont {Frauenheim}}, and\ \bibinfo
  {author} {\bibfnamefont {A.}~\bibnamefont {Gali}},\ }\href
  {https://doi.org/10.1103/PhysRevB.89.075203} {\bibfield  {journal} {\bibinfo
  {journal} {Phys. Rev. B}\ }\textbf {\bibinfo {volume} {89}},\ \bibinfo
  {pages} {075203} (\bibinfo {year} {2014})}\BibitemShut {NoStop}%
\bibitem [{\citenamefont {Neu}\ \emph {et~al.}(2011)\citenamefont {Neu},
  \citenamefont {Steinmetz}, \citenamefont {Riedrich-M\"{o}ller}, \citenamefont
  {Gsell}, \citenamefont {Fischer}, \citenamefont {Schreck},\ and\
  \citenamefont {Becher}}]{neu2011single}%
  \BibitemOpen
  \bibfield  {author} {\bibinfo {author} {\bibfnamefont {E.}~\bibnamefont
  {Neu}}, \bibinfo {author} {\bibfnamefont {D.}~\bibnamefont {Steinmetz}},
  \bibinfo {author} {\bibfnamefont {J.}~\bibnamefont {Riedrich-M\"{o}ller}},
  \bibinfo {author} {\bibfnamefont {S.}~\bibnamefont {Gsell}}, \bibinfo
  {author} {\bibfnamefont {M.}~\bibnamefont {Fischer}}, \bibinfo {author}
  {\bibfnamefont {M.}~\bibnamefont {Schreck}}, and\ \bibinfo {author}
  {\bibfnamefont {C.}~\bibnamefont {Becher}},\ }\href
  {https://doi.org/10.1088/1367-2630/13/2/025012} {\bibfield  {journal}
  {\bibinfo  {journal} {New J. Phys.}\ }\textbf {\bibinfo {volume} {13}},\
  \bibinfo {pages} {025012} (\bibinfo {year} {2011})}\BibitemShut {NoStop}%
\bibitem [{\citenamefont {H{\"a}u{\ss}ler}\ \emph {et~al.}(2017)\citenamefont
  {H{\"a}u{\ss}ler}, \citenamefont {Thiering}, \citenamefont {Dietrich},
  \citenamefont {Waasem}, \citenamefont {Teraji}, \citenamefont {Isoya},
  \citenamefont {Iwasaki}, \citenamefont {Hatano}, \citenamefont {Jelezko},
  \citenamefont {Gali},\ and\ \citenamefont
  {Kubanek}}]{haussler2017photoluminescence}%
  \BibitemOpen
  \bibfield  {author} {\bibinfo {author} {\bibfnamefont {S.}~\bibnamefont
  {H{\"a}u{\ss}ler}}, \bibinfo {author} {\bibfnamefont {G.}~\bibnamefont
  {Thiering}}, \bibinfo {author} {\bibfnamefont {A.}~\bibnamefont {Dietrich}},
  \bibinfo {author} {\bibfnamefont {N.}~\bibnamefont {Waasem}}, \bibinfo
  {author} {\bibfnamefont {T.}~\bibnamefont {Teraji}}, \bibinfo {author}
  {\bibfnamefont {J.}~\bibnamefont {Isoya}}, \bibinfo {author} {\bibfnamefont
  {T.}~\bibnamefont {Iwasaki}}, \bibinfo {author} {\bibfnamefont
  {M.}~\bibnamefont {Hatano}}, \bibinfo {author} {\bibfnamefont
  {F.}~\bibnamefont {Jelezko}}, \bibinfo {author} {\bibfnamefont
  {A.}~\bibnamefont {Gali}}, and\ \bibinfo {author} {\bibfnamefont
  {A.}~\bibnamefont {Kubanek}},\ }\href
  {https://doi.org/10.1088\%2F1367-2630\%2Faa73e5} {\bibfield  {journal}
  {\bibinfo  {journal} {New J. Phys.}\ }\textbf {\bibinfo {volume} {19}},\
  \bibinfo {pages} {063036} (\bibinfo {year} {2017})}\BibitemShut {NoStop}%
\bibitem [{\citenamefont {Ekimov}\ \emph {et~al.}(2015)\citenamefont {Ekimov},
  \citenamefont {Lyapin}, \citenamefont {Boldyrev}, \citenamefont {Kondrin},
  \citenamefont {Khmelnitskiy}, \citenamefont {Gavva}, \citenamefont
  {Kotereva},\ and\ \citenamefont {Popova}}]{Ekimov}%
  \BibitemOpen
  \bibfield  {author} {\bibinfo {author} {\bibfnamefont {E.~A.}\ \bibnamefont
  {Ekimov}}, \bibinfo {author} {\bibfnamefont {S.~G.}\ \bibnamefont {Lyapin}},
  \bibinfo {author} {\bibfnamefont {K.~N.}\ \bibnamefont {Boldyrev}}, \bibinfo
  {author} {\bibfnamefont {M.~V.}\ \bibnamefont {Kondrin}}, \bibinfo {author}
  {\bibfnamefont {R.}~\bibnamefont {Khmelnitskiy}}, \bibinfo {author}
  {\bibfnamefont {V.~A.}\ \bibnamefont {Gavva}}, \bibinfo {author}
  {\bibfnamefont {T.~V.}\ \bibnamefont {Kotereva}}, and\ \bibinfo {author}
  {\bibfnamefont {M.~N.}\ \bibnamefont {Popova}},\ }\href
  {https://doi.org/10.1134/S0021364015230034} {\bibfield  {journal} {\bibinfo
  {journal} {JETP Lett.}\ }\textbf {\bibinfo {volume} {102}},\ \bibinfo {pages}
  {701} (\bibinfo {year} {2015})}\BibitemShut {NoStop}%
\bibitem [{\citenamefont {Iwasaki}\ \emph {et~al.}(2015)\citenamefont
  {Iwasaki}, \citenamefont {Ishibashi}, \citenamefont {Miyamoto}, \citenamefont
  {Doi}, \citenamefont {Kobayashi}, \citenamefont {Miyazaki}, \citenamefont
  {Tahara}, \citenamefont {Jahnke}, \citenamefont {Rogers}, \citenamefont
  {Naydenov}, \citenamefont {Jelezko}, \citenamefont {Yamasaki}, \citenamefont
  {Nagamachi}, \citenamefont {Inubushi}, \citenamefont {Mizuochi},\ and\
  \citenamefont {Hatano}}]{Iwasaki}%
  \BibitemOpen
  \bibfield  {author} {\bibinfo {author} {\bibfnamefont {T.}~\bibnamefont
  {Iwasaki}}, \bibinfo {author} {\bibfnamefont {F.}~\bibnamefont {Ishibashi}},
  \bibinfo {author} {\bibfnamefont {Y.}~\bibnamefont {Miyamoto}}, \bibinfo
  {author} {\bibfnamefont {Y.}~\bibnamefont {Doi}}, \bibinfo {author}
  {\bibfnamefont {S.}~\bibnamefont {Kobayashi}}, \bibinfo {author}
  {\bibfnamefont {T.}~\bibnamefont {Miyazaki}}, \bibinfo {author}
  {\bibfnamefont {K.}~\bibnamefont {Tahara}}, \bibinfo {author} {\bibfnamefont
  {K.~D.}\ \bibnamefont {Jahnke}}, \bibinfo {author} {\bibfnamefont {L.~J.}\
  \bibnamefont {Rogers}}, \bibinfo {author} {\bibfnamefont {B.}~\bibnamefont
  {Naydenov}}, \bibinfo {author} {\bibfnamefont {F.}~\bibnamefont {Jelezko}},
  \bibinfo {author} {\bibfnamefont {S.}~\bibnamefont {Yamasaki}}, \bibinfo
  {author} {\bibfnamefont {S.}~\bibnamefont {Nagamachi}}, \bibinfo {author}
  {\bibfnamefont {T.}~\bibnamefont {Inubushi}}, \bibinfo {author}
  {\bibfnamefont {N.}~\bibnamefont {Mizuochi}}, and\ \bibinfo {author}
  {\bibfnamefont {M.}~\bibnamefont {Hatano}},\ }\href
  {https://doi.org/10.1038/srep12882} {\bibfield  {journal} {\bibinfo
  {journal} {Sci. Rep.}\ }\textbf {\bibinfo {volume} {5}},\ \bibinfo {pages}
  {12882} (\bibinfo {year} {2015})}\BibitemShut {NoStop}%
\bibitem [{\citenamefont {Ralchenko}\ \emph {et~al.}(2015)\citenamefont
  {Ralchenko}, \citenamefont {Sedov}, \citenamefont {Khomich}, \citenamefont
  {Krivobok}, \citenamefont {Nikolaev}, \citenamefont {Savin}, \citenamefont
  {Vlasov},\ and\ \citenamefont {Konov}}]{Ralchenko}%
  \BibitemOpen
  \bibfield  {author} {\bibinfo {author} {\bibfnamefont {V.~G.}\ \bibnamefont
  {Ralchenko}}, \bibinfo {author} {\bibfnamefont {V.~S.}\ \bibnamefont
  {Sedov}}, \bibinfo {author} {\bibfnamefont {A.~A.}\ \bibnamefont {Khomich}},
  \bibinfo {author} {\bibfnamefont {V.~S.}\ \bibnamefont {Krivobok}}, \bibinfo
  {author} {\bibfnamefont {S.~N.}\ \bibnamefont {Nikolaev}}, \bibinfo {author}
  {\bibfnamefont {S.}~\bibnamefont {Savin}}, \bibinfo {author} {\bibfnamefont
  {I.~I.}\ \bibnamefont {Vlasov}}, and\ \bibinfo {author} {\bibfnamefont
  {V.~I.}\ \bibnamefont {Konov}},\ }\href
  {https://doi.org/10.3103/S1068335615060020} {\bibfield  {journal} {\bibinfo
  {journal} {Bull. Leb. Phys. Inst.}\ }\textbf {\bibinfo {volume} {42}},\
  \bibinfo {pages} {165} (\bibinfo {year} {2015})}\BibitemShut {NoStop}%
\bibitem [{\citenamefont {Ekimov}\ \emph {et~al.}(2017)\citenamefont {Ekimov},
  \citenamefont {Krivobok}, \citenamefont {Lyapin}, \citenamefont {Sherin},
  \citenamefont {Gavva},\ and\ \citenamefont {Kondrin}}]{Ekimov2}%
  \BibitemOpen
  \bibfield  {author} {\bibinfo {author} {\bibfnamefont {E.~A.}\ \bibnamefont
  {Ekimov}}, \bibinfo {author} {\bibfnamefont {V.~S.}\ \bibnamefont
  {Krivobok}}, \bibinfo {author} {\bibfnamefont {S.~G.}\ \bibnamefont
  {Lyapin}}, \bibinfo {author} {\bibfnamefont {P.~S.}\ \bibnamefont {Sherin}},
  \bibinfo {author} {\bibfnamefont {V.~A.}\ \bibnamefont {Gavva}}, and\
  \bibinfo {author} {\bibfnamefont {M.~V.}\ \bibnamefont {Kondrin}},\ }\href
  {https://doi.org/10.1103/PhysRevB.95.094113} {\bibfield  {journal} {\bibinfo
  {journal} {Phys. Rev. B}\ }\textbf {\bibinfo {volume} {95}},\ \bibinfo
  {pages} {094113} (\bibinfo {year} {2017})}\BibitemShut {NoStop}%
\bibitem [{\citenamefont {Palyanov}\ \emph {et~al.}(2016)\citenamefont
  {Palyanov}, \citenamefont {Kupriyanov}, \citenamefont {Borzdov},
  \citenamefont {Khokhryakov},\ and\ \citenamefont {Surovtsev}}]{Palyanov}%
  \BibitemOpen
  \bibfield  {author} {\bibinfo {author} {\bibfnamefont {Y.~N.}\ \bibnamefont
  {Palyanov}}, \bibinfo {author} {\bibfnamefont {I.~N.}\ \bibnamefont
  {Kupriyanov}}, \bibinfo {author} {\bibfnamefont {Y.~M.}\ \bibnamefont
  {Borzdov}}, \bibinfo {author} {\bibfnamefont {A.~F.}\ \bibnamefont
  {Khokhryakov}}, and\ \bibinfo {author} {\bibfnamefont {N.~V.}\ \bibnamefont
  {Surovtsev}},\ }\href {https://doi.org/10.1021/acs.cgd.6b00481} {\bibfield
  {journal} {\bibinfo  {journal} {Cryst. Growth Des.}\ }\textbf {\bibinfo
  {volume} {16}},\ \bibinfo {pages} {3510} (\bibinfo {year}
  {2016})}\BibitemShut {NoStop}%
\bibitem [{\citenamefont {Palyanov}\ \emph {et~al.}(2015)\citenamefont
  {Palyanov}, \citenamefont {Kupriyanov}, \citenamefont {Borzdov},\ and\
  \citenamefont {Surovtsev}}]{Palyanov2015germanium}%
  \BibitemOpen
  \bibfield  {author} {\bibinfo {author} {\bibfnamefont {Y.~N.}\ \bibnamefont
  {Palyanov}}, \bibinfo {author} {\bibfnamefont {I.~N.}\ \bibnamefont
  {Kupriyanov}}, \bibinfo {author} {\bibfnamefont {Y.~M.}\ \bibnamefont
  {Borzdov}}, and\ \bibinfo {author} {\bibfnamefont {N.~V.}\ \bibnamefont
  {Surovtsev}},\ }\href {https://doi.org/10.1038/srep14789} {\bibfield
  {journal} {\bibinfo  {journal} {Sci. Rep.}\ }\textbf {\bibinfo {volume}
  {5}},\ \bibinfo {pages} {14789} (\bibinfo {year} {2015})}\BibitemShut
  {NoStop}%
\bibitem [{\citenamefont {Bhaskar}\ \emph {et~al.}(2017)\citenamefont
  {Bhaskar}, \citenamefont {Sukachev}, \citenamefont {Sipahigil}, \citenamefont
  {Evans}, \citenamefont {Burek}, \citenamefont {Nguyen}, \citenamefont
  {Rogers}, \citenamefont {Siyushev}, \citenamefont {Metsch}, \citenamefont
  {Park}, \citenamefont {Jelezko}, \citenamefont {Lon\ifmmode~\check{c}\else
  \v{c}\fi{}ar},\ and\ \citenamefont {Lukin}}]{Bhaskar}%
  \BibitemOpen
  \bibfield  {author} {\bibinfo {author} {\bibfnamefont {M.~K.}\ \bibnamefont
  {Bhaskar}}, \bibinfo {author} {\bibfnamefont {D.~D.}\ \bibnamefont
  {Sukachev}}, \bibinfo {author} {\bibfnamefont {A.}~\bibnamefont {Sipahigil}},
  \bibinfo {author} {\bibfnamefont {R.~E.}\ \bibnamefont {Evans}}, \bibinfo
  {author} {\bibfnamefont {M.~J.}\ \bibnamefont {Burek}}, \bibinfo {author}
  {\bibfnamefont {C.~T.}\ \bibnamefont {Nguyen}}, \bibinfo {author}
  {\bibfnamefont {L.~J.}\ \bibnamefont {Rogers}}, \bibinfo {author}
  {\bibfnamefont {P.}~\bibnamefont {Siyushev}}, \bibinfo {author}
  {\bibfnamefont {M.~H.}\ \bibnamefont {Metsch}}, \bibinfo {author}
  {\bibfnamefont {H.}~\bibnamefont {Park}}, \bibinfo {author} {\bibfnamefont
  {F.}~\bibnamefont {Jelezko}}, \bibinfo {author} {\bibfnamefont
  {M.}~\bibnamefont {Lon\ifmmode~\check{c}\else \v{c}\fi{}ar}}, and\ \bibinfo
  {author} {\bibfnamefont {M.~D.}\ \bibnamefont {Lukin}},\ }\href
  {https://doi.org/10.1103/PhysRevLett.118.223603} {\bibfield  {journal}
  {\bibinfo  {journal} {Phys. Rev. Lett.}\ }\textbf {\bibinfo {volume} {118}},\
  \bibinfo {pages} {223603} (\bibinfo {year} {2017})}\BibitemShut {NoStop}%
\bibitem [{\citenamefont {Bray}\ \emph {et~al.}(2018)\citenamefont {Bray},
  \citenamefont {Regan}, \citenamefont {Trycz}, \citenamefont {Previdi},
  \citenamefont {Seniutinas}, \citenamefont {Ganesan}, \citenamefont
  {Kianinia}, \citenamefont {Kim},\ and\ \citenamefont
  {Aharonovich}}]{bray2018single}%
  \BibitemOpen
  \bibfield  {author} {\bibinfo {author} {\bibfnamefont {K.}~\bibnamefont
  {Bray}}, \bibinfo {author} {\bibfnamefont {B.}~\bibnamefont {Regan}},
  \bibinfo {author} {\bibfnamefont {A.}~\bibnamefont {Trycz}}, \bibinfo
  {author} {\bibfnamefont {R.}~\bibnamefont {Previdi}}, \bibinfo {author}
  {\bibfnamefont {G.}~\bibnamefont {Seniutinas}}, \bibinfo {author}
  {\bibfnamefont {K.}~\bibnamefont {Ganesan}}, \bibinfo {author} {\bibfnamefont
  {M.}~\bibnamefont {Kianinia}}, \bibinfo {author} {\bibfnamefont
  {S.}~\bibnamefont {Kim}}, and\ \bibinfo {author} {\bibfnamefont
  {I.}~\bibnamefont {Aharonovich}},\ }\href
  {https://doi.org/10.1021/acsphotonics.8b00930} {\bibfield  {journal}
  {\bibinfo  {journal} {ACS Photonics}\ }\textbf {\bibinfo {volume} {5}},\
  \bibinfo {pages} {4817} (\bibinfo {year} {2018})}\BibitemShut {NoStop}%
\bibitem [{\citenamefont {Iwasaki}\ \emph {et~al.}(2017)\citenamefont
  {Iwasaki}, \citenamefont {Miyamoto}, \citenamefont {Taniguchi}, \citenamefont
  {Siyushev}, \citenamefont {Metsch}, \citenamefont {Jelezko},\ and\
  \citenamefont {Hatano}}]{Iwasaki2}%
  \BibitemOpen
  \bibfield  {author} {\bibinfo {author} {\bibfnamefont {T.}~\bibnamefont
  {Iwasaki}}, \bibinfo {author} {\bibfnamefont {Y.}~\bibnamefont {Miyamoto}},
  \bibinfo {author} {\bibfnamefont {T.}~\bibnamefont {Taniguchi}}, \bibinfo
  {author} {\bibfnamefont {P.}~\bibnamefont {Siyushev}}, \bibinfo {author}
  {\bibfnamefont {M.~H.}\ \bibnamefont {Metsch}}, \bibinfo {author}
  {\bibfnamefont {F.}~\bibnamefont {Jelezko}}, and\ \bibinfo {author}
  {\bibfnamefont {M.}~\bibnamefont {Hatano}},\ }\href
  {https://doi.org/10.1103/PhysRevLett.119.253601} {\bibfield  {journal}
  {\bibinfo  {journal} {Phys. Rev. Lett.}\ }\textbf {\bibinfo {volume} {119}},\
  \bibinfo {pages} {253601} (\bibinfo {year} {2017})}\BibitemShut {NoStop}%
\bibitem [{\citenamefont {Ekimov}\ \emph {et~al.}(2018)\citenamefont {Ekimov},
  \citenamefont {Lyapin},\ and\ \citenamefont {Kondrin}}]{ekimov2018tin}%
  \BibitemOpen
  \bibfield  {author} {\bibinfo {author} {\bibfnamefont {E.}~\bibnamefont
  {Ekimov}}, \bibinfo {author} {\bibfnamefont {S.}~\bibnamefont {Lyapin}}, and\
  \bibinfo {author} {\bibfnamefont {M.}~\bibnamefont {Kondrin}},\ }\href
  {https://doi.org/https://doi.org/10.1016/j.diamond.2018.06.014} {\bibfield
  {journal} {\bibinfo  {journal} {Diam. Relat. Mater.}\ }\textbf {\bibinfo
  {volume} {87}},\ \bibinfo {pages} {223} (\bibinfo {year} {2018})}\BibitemShut
  {NoStop}%
\bibitem [{\citenamefont {Ditalia~Tchernij}\ \emph {et~al.}(2017)\citenamefont
  {Ditalia~Tchernij}, \citenamefont {Herzig}, \citenamefont {Forneris},
  \citenamefont {K\"upper}, \citenamefont {Pezzagna}, \citenamefont {Traina},
  \citenamefont {Moreva}, \citenamefont {Degiovanni}, \citenamefont {Brida},
  \citenamefont {Skukan}, \citenamefont {Genovese}, \citenamefont
  {Jak\ifmmode~\breve{s}\else \u{s}\fi{}i\'c}, \citenamefont {Meijer},\ and\
  \citenamefont {Olivero}}]{tchernij2017single}%
  \BibitemOpen
  \bibfield  {author} {\bibinfo {author} {\bibfnamefont {S.}~\bibnamefont
  {Ditalia~Tchernij}}, \bibinfo {author} {\bibfnamefont {T.}~\bibnamefont
  {Herzig}}, \bibinfo {author} {\bibfnamefont {J.}~\bibnamefont {Forneris}},
  \bibinfo {author} {\bibfnamefont {J.}~\bibnamefont {K\"upper}}, \bibinfo
  {author} {\bibfnamefont {S.}~\bibnamefont {Pezzagna}}, \bibinfo {author}
  {\bibfnamefont {P.}~\bibnamefont {Traina}}, \bibinfo {author} {\bibfnamefont
  {E.}~\bibnamefont {Moreva}}, \bibinfo {author} {\bibfnamefont {I.~P.}\
  \bibnamefont {Degiovanni}}, \bibinfo {author} {\bibfnamefont
  {G.}~\bibnamefont {Brida}}, \bibinfo {author} {\bibfnamefont
  {N.}~\bibnamefont {Skukan}}, \bibinfo {author} {\bibfnamefont
  {M.}~\bibnamefont {Genovese}}, \bibinfo {author} {\bibfnamefont
  {M.}~\bibnamefont {Jak\ifmmode~\breve{s}\else \u{s}\fi{}i\'c}}, \bibinfo
  {author} {\bibfnamefont {J.}~\bibnamefont {Meijer}}, and\ \bibinfo {author}
  {\bibfnamefont {P.}~\bibnamefont {Olivero}},\ }\href
  {https://doi.org/10.1021/acsphotonics.7b00904} {\bibfield  {journal}
  {\bibinfo  {journal} {ACS Photonics}\ }\textbf {\bibinfo {volume} {4}},\
  \bibinfo {pages} {2580} (\bibinfo {year} {2017})}\BibitemShut {NoStop}%
\bibitem [{\citenamefont {Palyanov}\ \emph {et~al.}(2019)\citenamefont
  {Palyanov}, \citenamefont {Kupriyanov},\ and\ \citenamefont
  {Borzdov}}]{palyanov2019high}%
  \BibitemOpen
  \bibfield  {author} {\bibinfo {author} {\bibfnamefont {Y.~N.}\ \bibnamefont
  {Palyanov}}, \bibinfo {author} {\bibfnamefont {I.~N.}\ \bibnamefont
  {Kupriyanov}}, and\ \bibinfo {author} {\bibfnamefont {Y.~M.}\ \bibnamefont
  {Borzdov}},\ }\href
  {https://doi.org/https://doi.org/10.1016/j.carbon.2018.11.084} {\bibfield
  {journal} {\bibinfo  {journal} {Carbon}\ }\textbf {\bibinfo {volume} {143}},\
  \bibinfo {pages} {769} (\bibinfo {year} {2019})}\BibitemShut {NoStop}%
\bibitem [{\citenamefont {Alkahtani}\ \emph {et~al.}(2018)\citenamefont
  {Alkahtani}, \citenamefont {Cojocaru}, \citenamefont {Liu}, \citenamefont
  {Herzig}, \citenamefont {Meijer}, \citenamefont {K{\"u}pper}, \citenamefont
  {L{\"u}hmann}, \citenamefont {Akimov},\ and\ \citenamefont
  {Hemmer}}]{Alkahtani2018tin}%
  \BibitemOpen
  \bibfield  {author} {\bibinfo {author} {\bibfnamefont {M.}~\bibnamefont
  {Alkahtani}}, \bibinfo {author} {\bibfnamefont {I.}~\bibnamefont {Cojocaru}},
  \bibinfo {author} {\bibfnamefont {X.}~\bibnamefont {Liu}}, \bibinfo {author}
  {\bibfnamefont {T.}~\bibnamefont {Herzig}}, \bibinfo {author} {\bibfnamefont
  {J.}~\bibnamefont {Meijer}}, \bibinfo {author} {\bibfnamefont
  {J.}~\bibnamefont {K{\"u}pper}}, \bibinfo {author} {\bibfnamefont
  {T.}~\bibnamefont {L{\"u}hmann}}, \bibinfo {author} {\bibfnamefont {A.~V.}\
  \bibnamefont {Akimov}}, and\ \bibinfo {author} {\bibfnamefont {P.~R.}\
  \bibnamefont {Hemmer}},\ }\href {https://doi.org/10.1063/1.5037053}
  {\bibfield  {journal} {\bibinfo  {journal} {Appl. Phys. Lett.}\ }\textbf
  {\bibinfo {volume} {112}},\ \bibinfo {pages} {241902} (\bibinfo {year}
  {2018})}\BibitemShut {NoStop}%
\bibitem [{\citenamefont {Rugar}\ \emph {et~al.}(2019)\citenamefont {Rugar},
  \citenamefont {Dory}, \citenamefont {Sun},\ and\ \citenamefont {Vu\ifmmode
  \check{c}\else \v{c}\fi{}kovi\ifmmode~\acute{c}\else
  \'{c}\fi{}}}]{rugar2019char}%
  \BibitemOpen
  \bibfield  {author} {\bibinfo {author} {\bibfnamefont {A.~E.}\ \bibnamefont
  {Rugar}}, \bibinfo {author} {\bibfnamefont {C.}~\bibnamefont {Dory}},
  \bibinfo {author} {\bibfnamefont {S.}~\bibnamefont {Sun}}, and\ \bibinfo
  {author} {\bibfnamefont {J.}~\bibnamefont {Vu\ifmmode \check{c}\else
  \v{c}\fi{}kovi\ifmmode~\acute{c}\else \'{c}\fi{}}},\ }\href
  {https://doi.org/10.1103/PhysRevB.99.205417} {\bibfield  {journal} {\bibinfo
  {journal} {Phys. Rev. B}\ }\textbf {\bibinfo {volume} {99}},\ \bibinfo
  {pages} {205417} (\bibinfo {year} {2019})}\BibitemShut {NoStop}%
\bibitem [{\citenamefont {Wahl}\ \emph {et~al.}(2020)\citenamefont {Wahl},
  \citenamefont {Correia}, \citenamefont {Villarreal}, \citenamefont
  {Bourgeois}, \citenamefont {Gulka}, \citenamefont {Nesl{\'a}dek},
  \citenamefont {Vantomme},\ and\ \citenamefont {Pereira}}]{wahl2020direct}%
  \BibitemOpen
  \bibfield  {author} {\bibinfo {author} {\bibfnamefont {U.}~\bibnamefont
  {Wahl}}, \bibinfo {author} {\bibfnamefont {J.}~\bibnamefont {Correia}},
  \bibinfo {author} {\bibfnamefont {R.}~\bibnamefont {Villarreal}}, \bibinfo
  {author} {\bibfnamefont {E.}~\bibnamefont {Bourgeois}}, \bibinfo {author}
  {\bibfnamefont {M.}~\bibnamefont {Gulka}}, \bibinfo {author} {\bibfnamefont
  {M.}~\bibnamefont {Nesl{\'a}dek}}, \bibinfo {author} {\bibfnamefont
  {A.}~\bibnamefont {Vantomme}}, and\ \bibinfo {author} {\bibfnamefont
  {L.}~\bibnamefont {Pereira}},\ }\href
  {https://link.aps.org/doi/10.1103/PhysRevLett.125.045301} {\bibfield
  {journal} {\bibinfo  {journal} {Physical Review Letters}\ }\textbf {\bibinfo
  {volume} {125}},\ \bibinfo {pages} {045301} (\bibinfo {year}
  {2020})}\BibitemShut {NoStop}%
\bibitem [{\citenamefont {Fukuta}\ \emph {et~al.}(2021)\citenamefont {Fukuta},
  \citenamefont {Murakami}, \citenamefont {Ohfuji}, \citenamefont {Shinmei},
  \citenamefont {Irifune},\ and\ \citenamefont {Ishikawa}}]{fukuta2021sn}%
  \BibitemOpen
  \bibfield  {author} {\bibinfo {author} {\bibfnamefont {R.}~\bibnamefont
  {Fukuta}}, \bibinfo {author} {\bibfnamefont {Y.}~\bibnamefont {Murakami}},
  \bibinfo {author} {\bibfnamefont {H.}~\bibnamefont {Ohfuji}}, \bibinfo
  {author} {\bibfnamefont {T.}~\bibnamefont {Shinmei}}, \bibinfo {author}
  {\bibfnamefont {T.}~\bibnamefont {Irifune}}, and\ \bibinfo {author}
  {\bibfnamefont {F.}~\bibnamefont {Ishikawa}},\ }\href
  {https://doi.org/10.35848/1347-4065/abdc31} {\bibfield  {journal} {\bibinfo
  {journal} {Japanese Journal of Applied Physics}\ }\textbf {\bibinfo {volume}
  {60}},\ \bibinfo {pages} {035501} (\bibinfo {year} {2021})}\BibitemShut
  {NoStop}%
\bibitem [{\citenamefont {Trusheim}\ \emph {et~al.}(2019)\citenamefont
  {Trusheim}, \citenamefont {Wan}, \citenamefont {Chen}, \citenamefont
  {Ciccarino}, \citenamefont {Flick}, \citenamefont {Sundararaman},
  \citenamefont {Malladi}, \citenamefont {Bersin}, \citenamefont {Walsh},
  \citenamefont {Lienhard}, \citenamefont {Bakhru}, \citenamefont {Narang},\
  and\ \citenamefont {Englund}}]{Trusheim}%
  \BibitemOpen
  \bibfield  {author} {\bibinfo {author} {\bibfnamefont {M.~E.}\ \bibnamefont
  {Trusheim}}, \bibinfo {author} {\bibfnamefont {N.~H.}\ \bibnamefont {Wan}},
  \bibinfo {author} {\bibfnamefont {K.~C.}\ \bibnamefont {Chen}}, \bibinfo
  {author} {\bibfnamefont {C.~J.}\ \bibnamefont {Ciccarino}}, \bibinfo {author}
  {\bibfnamefont {J.}~\bibnamefont {Flick}}, \bibinfo {author} {\bibfnamefont
  {R.}~\bibnamefont {Sundararaman}}, \bibinfo {author} {\bibfnamefont
  {G.}~\bibnamefont {Malladi}}, \bibinfo {author} {\bibfnamefont
  {E.}~\bibnamefont {Bersin}}, \bibinfo {author} {\bibfnamefont
  {M.}~\bibnamefont {Walsh}}, \bibinfo {author} {\bibfnamefont
  {B.}~\bibnamefont {Lienhard}}, \bibinfo {author} {\bibfnamefont
  {H.}~\bibnamefont {Bakhru}}, \bibinfo {author} {\bibfnamefont
  {P.}~\bibnamefont {Narang}}, and\ \bibinfo {author} {\bibfnamefont
  {D.}~\bibnamefont {Englund}},\ }\href
  {https://doi.org/10.1103/PhysRevB.99.075430} {\bibfield  {journal} {\bibinfo
  {journal} {Phys. Rev. B}\ }\textbf {\bibinfo {volume} {99}},\ \bibinfo
  {pages} {075430} (\bibinfo {year} {2019})}\BibitemShut {NoStop}%
\bibitem [{\citenamefont {Ditalia~Tchernij}\ \emph {et~al.}(2018)\citenamefont
  {Ditalia~Tchernij}, \citenamefont {L\"{u}hmann}, \citenamefont {Herzig},
  \citenamefont {K\"{u}pper}, \citenamefont {Damin}, \citenamefont
  {Santonocito}, \citenamefont {Signorile}, \citenamefont {Traina},
  \citenamefont {Moreva}, \citenamefont {Celegato}, \citenamefont {Pezzagna},
  \citenamefont {Degiovanni}, \citenamefont {Olivero}, \citenamefont
  {Jak\ifmmode~\breve{s}\else \u{s}\fi{}i\'c}, \citenamefont {Meijer},
  \citenamefont {Genovese},\ and\ \citenamefont
  {Forneris}}]{tchernij2018single}%
  \BibitemOpen
  \bibfield  {author} {\bibinfo {author} {\bibfnamefont {S.}~\bibnamefont
  {Ditalia~Tchernij}}, \bibinfo {author} {\bibfnamefont {T.}~\bibnamefont
  {L\"{u}hmann}}, \bibinfo {author} {\bibfnamefont {T.}~\bibnamefont {Herzig}},
  \bibinfo {author} {\bibfnamefont {J.}~\bibnamefont {K\"{u}pper}}, \bibinfo
  {author} {\bibfnamefont {A.}~\bibnamefont {Damin}}, \bibinfo {author}
  {\bibfnamefont {S.}~\bibnamefont {Santonocito}}, \bibinfo {author}
  {\bibfnamefont {M.}~\bibnamefont {Signorile}}, \bibinfo {author}
  {\bibfnamefont {P.}~\bibnamefont {Traina}}, \bibinfo {author} {\bibfnamefont
  {E.}~\bibnamefont {Moreva}}, \bibinfo {author} {\bibfnamefont
  {F.}~\bibnamefont {Celegato}}, \bibinfo {author} {\bibfnamefont
  {S.}~\bibnamefont {Pezzagna}}, \bibinfo {author} {\bibfnamefont {I.~P.}\
  \bibnamefont {Degiovanni}}, \bibinfo {author} {\bibfnamefont
  {P.}~\bibnamefont {Olivero}}, \bibinfo {author} {\bibfnamefont
  {M.}~\bibnamefont {Jak\ifmmode~\breve{s}\else \u{s}\fi{}i\'c}}, \bibinfo
  {author} {\bibfnamefont {J.}~\bibnamefont {Meijer}}, \bibinfo {author}
  {\bibfnamefont {P.~M.}\ \bibnamefont {Genovese}}, and\ \bibinfo {author}
  {\bibfnamefont {J.}~\bibnamefont {Forneris}},\ }\href
  {https://doi.org/10.1021/acsphotonics.8b01013} {\bibfield  {journal}
  {\bibinfo  {journal} {ACS Photonics}\ }\textbf {\bibinfo {volume} {5}},\
  \bibinfo {pages} {4864} (\bibinfo {year} {2018})}\BibitemShut {NoStop}%
\bibitem [{\citenamefont {Ekimov}\ \emph {et~al.}(2019)\citenamefont {Ekimov},
  \citenamefont {Kondrin}, \citenamefont {Krivobok}, \citenamefont {Khomich},
  \citenamefont {Vlasov}, \citenamefont {Khmelnitskiy}, \citenamefont
  {Iwasaki},\ and\ \citenamefont {Hatano}}]{ekimov2019effect}%
  \BibitemOpen
  \bibfield  {author} {\bibinfo {author} {\bibfnamefont {E.}~\bibnamefont
  {Ekimov}}, \bibinfo {author} {\bibfnamefont {M.}~\bibnamefont {Kondrin}},
  \bibinfo {author} {\bibfnamefont {V.}~\bibnamefont {Krivobok}}, \bibinfo
  {author} {\bibfnamefont {A.}~\bibnamefont {Khomich}}, \bibinfo {author}
  {\bibfnamefont {I.}~\bibnamefont {Vlasov}}, \bibinfo {author} {\bibfnamefont
  {R.}~\bibnamefont {Khmelnitskiy}}, \bibinfo {author} {\bibfnamefont
  {T.}~\bibnamefont {Iwasaki}}, and\ \bibinfo {author} {\bibfnamefont
  {M.}~\bibnamefont {Hatano}},\ }\href
  {https://doi.org/https://doi.org/10.1016/j.diamond.2019.01.029} {\bibfield
  {journal} {\bibinfo  {journal} {Diam. Relat. Mater.}\ }\textbf {\bibinfo
  {volume} {93}},\ \bibinfo {pages} {75} (\bibinfo {year} {2019})}\BibitemShut
  {NoStop}%
\bibitem [{\citenamefont {Thiering$~$}\ and\ \citenamefont
  {Gali}(2018)}]{Gali3}%
  \BibitemOpen
  \bibfield  {author} {\bibinfo {author} {\bibfnamefont {G.}~\bibnamefont
  {Thiering$~$}}and\ \bibinfo {author} {\bibfnamefont {A.}~\bibnamefont
  {Gali}},\ }\href {https://doi.org/10.1103/PhysRevX.8.021063} {\bibfield
  {journal} {\bibinfo  {journal} {Phys. Rev. X}\ }\textbf {\bibinfo {volume}
  {8}},\ \bibinfo {pages} {021063} (\bibinfo {year} {2018})}\BibitemShut
  {NoStop}%
\bibitem [{\citenamefont {Gali$~$}\ and\ \citenamefont
  {Maze}(2013)}]{gali2013ab}%
  \BibitemOpen
  \bibfield  {author} {\bibinfo {author} {\bibfnamefont {A.}~\bibnamefont
  {Gali$~$}}and\ \bibinfo {author} {\bibfnamefont {J.~R.}\ \bibnamefont
  {Maze}},\ }\href {https://doi.org/10.1103/PhysRevB.88.235205} {\bibfield
  {journal} {\bibinfo  {journal} {Phys. Rev. B}\ }\textbf {\bibinfo {volume}
  {88}},\ \bibinfo {pages} {235205} (\bibinfo {year} {2013})}\BibitemShut
  {NoStop}%
\bibitem [{\citenamefont {Edmonds}\ \emph {et~al.}(2008)\citenamefont
  {Edmonds}, \citenamefont {Newton}, \citenamefont {Martineau}, \citenamefont
  {Twitchen},\ and\ \citenamefont {Williams}}]{edmonds2008electron}%
  \BibitemOpen
  \bibfield  {author} {\bibinfo {author} {\bibfnamefont {A.~M.}\ \bibnamefont
  {Edmonds}}, \bibinfo {author} {\bibfnamefont {M.~E.}\ \bibnamefont {Newton}},
  \bibinfo {author} {\bibfnamefont {P.~M.}\ \bibnamefont {Martineau}}, \bibinfo
  {author} {\bibfnamefont {D.~J.}\ \bibnamefont {Twitchen}}, and\ \bibinfo
  {author} {\bibfnamefont {S.~D.}\ \bibnamefont {Williams}},\ }\href
  {https://doi.org/10.1103/PhysRevB.77.245205} {\bibfield  {journal} {\bibinfo
  {journal} {Phys. Rev. B}\ }\textbf {\bibinfo {volume} {77}},\ \bibinfo
  {pages} {245205} (\bibinfo {year} {2008})}\BibitemShut {NoStop}%
\bibitem [{\citenamefont {Madelung}(1991)}]{madelung1991semiconductors}%
  \BibitemOpen
  \bibfield  {author} {\bibinfo {author} {\bibfnamefont {O.}~\bibnamefont
  {Madelung}},\ }\href {https://books.google.com/books?id=KbbvAAAAMAAJ} {\emph
  {\bibinfo {title} {Semiconductors: group IV elements and III-V compounds}}},\
  Data in science and technology\ (\bibinfo  {publisher} {Springer-Verlag},\
  \bibinfo {year} {1991})\BibitemShut {NoStop}%
\bibitem [{\citenamefont {Lany$~$}\ and\ \citenamefont {Zunger}(2008)}]{Lany}%
  \BibitemOpen
  \bibfield  {author} {\bibinfo {author} {\bibfnamefont {S.}~\bibnamefont
  {Lany$~$}}and\ \bibinfo {author} {\bibfnamefont {A.}~\bibnamefont {Zunger}},\
  }\href {https://doi.org/10.1103/PhysRevB.78.235104} {\bibfield  {journal}
  {\bibinfo  {journal} {Phys. Rev. B}\ }\textbf {\bibinfo {volume} {78}},\
  \bibinfo {pages} {235104} (\bibinfo {year} {2008})}\BibitemShut {NoStop}%
\bibitem [{\citenamefont {Nizovtsev}\ \emph {et~al.}(2020)\citenamefont
  {Nizovtsev}, \citenamefont {Kilin}, \citenamefont {Pushkarchuk},
  \citenamefont {Kuten}, \citenamefont {Poklonski}, \citenamefont {Michels},
  \citenamefont {Lyakhov},\ and\ \citenamefont {Jelezko}}]{Nizovtsev2020}%
  \BibitemOpen
  \bibfield  {author} {\bibinfo {author} {\bibfnamefont {A.~P.}\ \bibnamefont
  {Nizovtsev}}, \bibinfo {author} {\bibfnamefont {S.~Y.}\ \bibnamefont
  {Kilin}}, \bibinfo {author} {\bibfnamefont {A.~L.}\ \bibnamefont
  {Pushkarchuk}}, \bibinfo {author} {\bibfnamefont {S.~A.}\ \bibnamefont
  {Kuten}}, \bibinfo {author} {\bibfnamefont {N.~A.}\ \bibnamefont
  {Poklonski}}, \bibinfo {author} {\bibfnamefont {D.}~\bibnamefont {Michels}},
  \bibinfo {author} {\bibfnamefont {D.}~\bibnamefont {Lyakhov}}, and\ \bibinfo
  {author} {\bibfnamefont {F.}~\bibnamefont {Jelezko}},\ }\href
  {https://doi.org/10.1134/S1063782620120271} {\bibfield  {journal} {\bibinfo
  {journal} {Semiconductors}\ }\textbf {\bibinfo {volume} {54}},\ \bibinfo
  {pages} {1685} (\bibinfo {year} {2020})}\BibitemShut {NoStop}%
\bibitem [{\citenamefont {Ciccarino}\ \emph {et~al.}(2020)\citenamefont
  {Ciccarino}, \citenamefont {Flick}, \citenamefont {Harris}, \citenamefont
  {Trusheim}, \citenamefont {Englund},\ and\ \citenamefont
  {Narang}}]{ciccarino2020strong}%
  \BibitemOpen
  \bibfield  {author} {\bibinfo {author} {\bibfnamefont {C.~J.}\ \bibnamefont
  {Ciccarino}}, \bibinfo {author} {\bibfnamefont {J.}~\bibnamefont {Flick}},
  \bibinfo {author} {\bibfnamefont {I.~B.}\ \bibnamefont {Harris}}, \bibinfo
  {author} {\bibfnamefont {M.~E.}\ \bibnamefont {Trusheim}}, \bibinfo {author}
  {\bibfnamefont {D.~R.}\ \bibnamefont {Englund}}, and\ \bibinfo {author}
  {\bibfnamefont {P.}~\bibnamefont {Narang}},\ }\href
  {https://doi.org/10.1038/s41535-020-00281-7} {\bibfield  {journal} {\bibinfo
  {journal} {npj Quantum Materials}\ }\textbf {\bibinfo {volume} {5}},\
  \bibinfo {pages} {75} (\bibinfo {year} {2020})}\BibitemShut {NoStop}%
\bibitem [{\citenamefont {Chen}\ \emph {et~al.}(2020)\citenamefont {Chen},
  \citenamefont {Hu}, \citenamefont {Stanton}, \citenamefont {Hill},
  \citenamefont {Cheng},\ and\ \citenamefont {Zhang}}]{chen2020decoherence}%
  \BibitemOpen
  \bibfield  {author} {\bibinfo {author} {\bibfnamefont {J.}~\bibnamefont
  {Chen}}, \bibinfo {author} {\bibfnamefont {C.}~\bibnamefont {Hu}}, \bibinfo
  {author} {\bibfnamefont {J.~F.}\ \bibnamefont {Stanton}}, \bibinfo {author}
  {\bibfnamefont {S.}~\bibnamefont {Hill}}, \bibinfo {author} {\bibfnamefont
  {H.-P.}\ \bibnamefont {Cheng}}, and\ \bibinfo {author} {\bibfnamefont
  {X.-G.}\ \bibnamefont {Zhang}},\ }\bibfield  {booktitle} {\emph {\bibinfo
  {booktitle} {The Journal of Physical Chemistry Letters}},\ }\href
  {https://doi.org/10.1021/acs.jpclett.0c00193} {\bibfield  {journal} {\bibinfo
   {journal} {The Journal of Physical Chemistry Letters}\ }\textbf {\bibinfo
  {volume} {11}},\ \bibinfo {pages} {2074} (\bibinfo {year}
  {2020})}\BibitemShut {NoStop}%
\bibitem [{\citenamefont {Bloembergen}(1949)}]{BLOEMBERGEN1949on}%
  \BibitemOpen
  \bibfield  {author} {\bibinfo {author} {\bibfnamefont {N.}~\bibnamefont
  {Bloembergen}},\ }\href
  {https://doi.org/https://doi.org/10.1016/0031-8914(49)90114-7} {\bibfield
  {journal} {\bibinfo  {journal} {Physica}\ }\textbf {\bibinfo {volume} {15}},\
  \bibinfo {pages} {386} (\bibinfo {year} {1949})}\BibitemShut {NoStop}%
\bibitem [{\citenamefont {Towns}\ \emph {et~al.}(2014)\citenamefont {Towns},
  \citenamefont {Cockerill}, \citenamefont {Dahan}, \citenamefont {Foster},
  \citenamefont {Gaither}, \citenamefont {Grimshaw}, \citenamefont {Hazlewood},
  \citenamefont {Lathrop}, \citenamefont {Lifka}, \citenamefont {Peterson},
  \citenamefont {Roskies}, \citenamefont {Scott},\ and\ \citenamefont
  {Wilkins-Diehr}}]{Towns}%
  \BibitemOpen
  \bibfield  {author} {\bibinfo {author} {\bibfnamefont {J.}~\bibnamefont
  {Towns}}, \bibinfo {author} {\bibfnamefont {T.}~\bibnamefont {Cockerill}},
  \bibinfo {author} {\bibfnamefont {M.}~\bibnamefont {Dahan}}, \bibinfo
  {author} {\bibfnamefont {I.}~\bibnamefont {Foster}}, \bibinfo {author}
  {\bibfnamefont {K.}~\bibnamefont {Gaither}}, \bibinfo {author} {\bibfnamefont
  {A.}~\bibnamefont {Grimshaw}}, \bibinfo {author} {\bibfnamefont
  {V.}~\bibnamefont {Hazlewood}}, \bibinfo {author} {\bibfnamefont
  {S.}~\bibnamefont {Lathrop}}, \bibinfo {author} {\bibfnamefont
  {D.}~\bibnamefont {Lifka}}, \bibinfo {author} {\bibfnamefont {G.~D.}\
  \bibnamefont {Peterson}}, \bibinfo {author} {\bibfnamefont {R.}~\bibnamefont
  {Roskies}}, \bibinfo {author} {\bibfnamefont {J.~R.}\ \bibnamefont {Scott}},
  and\ \bibinfo {author} {\bibfnamefont {N.}~\bibnamefont {Wilkins-Diehr}},\
  }\href {https://doi.org/10.1109/MCSE.2014.80} {\bibfield  {journal} {\bibinfo
   {journal} {Computing in Science \& Engineering}\ }\textbf {\bibinfo {volume}
  {16}},\ \bibinfo {pages} {62} (\bibinfo {year} {2014})}\BibitemShut {NoStop}%
\end{thebibliography}%

\end{document}